\documentclass[fleqn,usenatbib]{mnras}

\usepackage{newtxtext,newtxmath}
\usepackage{breqn}

\usepackage[T1]{fontenc}
\usepackage{ae,aecompl}
\usepackage[dvipsnames]{xcolor}
\usepackage[toc,page]{appendix}
\usepackage{pdflscape}
\usepackage{longtable}
\usepackage{booktabs}

\usepackage{graphicx} 
\usepackage{amsmath} 
\usepackage{rotating}

\title[Type I X-ray Bursts During Eclipses]{Type I X-ray Burst Emission Reflected into the Eclipses of EXO 0748$-$676}

\author[A. H. Knight et al.]{
Amy H. Knight,$^{1,2}$\thanks{E-mail: amy.h.knight@durham.ac.uk}
Jakob van den Eijnden,$^{3, 2}$
Adam Ingram,$^{4,2}$ 
James H. Matthews,$^{2}$
Sara E. Motta,$^{5}$\and
\ Matthew Middleton,$^{6}$ 
Giulio C. Mancuso,$^{7,8}$
Douglas J. K. Buisson,$^{9}$
Diego Altamirano,$^{6}$
Rob Fender$^{2}$\and
\ and Timothy P. Roberts$^{1}$ \\
\\
$^{1}$ Centre for Extragalactic Astronomy, Department of Physics, Durham University, South Road, Durham DH1 3LE, UK \\
$^{2}$Department of Physics, Astrophysics, University of Oxford, Denys Wilkinson Building, Keble Road, Oxford, OX1 3RH, UK\\
$^{3}$Department of Physics, Gibbet Hill Road, University of Warwick, Coventry, CV4 7AL, United Kingdom\\
$^{4}$School of Mathematics, Statistics, and Physics, Newcastle University, Newcastle upon Tyne, NE1 7RU, UK\\
$^{5}$INAF, Osservatorio Astronomico di Brera, Via E. Bianchi 46, I-23807 Merate (LC), Italy\\
$^{6}$School of Physics and Astronomy, University of Southampton, Highfield, Southampton, SO17 1BJ, UK\\
$^{7}$Instituto Argentino de Radioastronom\'{i}a (CCT-La Plata, CONICET; CICPBA), C.C. No. 5, 1894 Villa Elisa, Argentina\\
$^{8}$Facultad de Ciencias Astron\'{o}micas y Geof\'{i}sicas, Universidad Nacional de La Plata, Paseo del Bosque, 1900 La Plata, Argentina\\
$^{9}$Independent Researcher\\
}

\date{Accepted XXX. Received YYY; in original form ZZZ}

\pubyear{2024}

\begin{document}
\label{firstpage}
\pagerange{\pageref{firstpage}--\pageref{lastpage}}
\maketitle

\begin{abstract}
The neutron star X-ray binary, EXO 0748--676, was observed regularly by the Rossi X-ray Timing Explorer (\textit{RXTE}) and \textit{XMM-Newton} during its first detected outburst (1985 - 2008). These observations captured hundreds of asymmetric, energy-dependent X-ray eclipses, influenced by the ongoing ablation of the companion star and numerous Type I thermonuclear X-ray bursts. Here, we present the light curves of 22 Type I X-ray bursts observed by \textit{RXTE} that coincide, fully or partially, with an X-ray eclipse. We identify nine instances where the burst occurs entirely within totality, seven bursts split across an egress, and six cases interrupted by an ingress. All in-eclipse bursts and split bursts occurred while the source was in the hard spectral state. We establish that we are not observing direct burst emission during eclipses since the companion star and the ablated outflow entirely obscure our view of the X-ray emitting region. We determine that the reflected flux from the outer accretion disc, even if maximally flared, is insufficient to explain all observations of in-eclipse X-ray bursts and instead explore scenarios whereby the emission arising from the X-ray bursts is scattered, either by a burst-induced rise in $N_{\rm{H}}$ that provides extra material, an accretion disc wind or the ablated outflow, into our line of sight. However, the rarity of a burst and eclipse overlap makes it challenging to determine their origin.
\end{abstract}




\section{Introduction}
Neutron star (NS) low-mass X-ray binaries (LMXBs) are systems consisting of a NS accreting material from a companion star via Roche lobe overflow (RLOF), which then forms an accretion disc around it (\citealt{Tauris2006}; see \citealt{Bahramian2022} for a review). While actively accreting, some of the material accumulates on the surface of the NS. Upon reaching sufficient temperature and density, the material layer built up on the NS surface ignites, causing thermonuclear runaway and resulting in observable bursts of X-ray emission (e.g. \citealt{Lewin1993, Strohmayer2006} and references therein). These Type I X-ray bursts (also called thermonuclear bursts or X-ray bursts) manifest as rapid and sudden increases in the observed X-ray flux, peaking much brighter than the level of persistent X-ray emission (e.g. \citealt{ Galloway2008, Galloway2020}) and are one of the few observable events to uniquely identify the compact object in accreting systems as a NS rather than a black hole.

Type I X-ray bursts result from unstable thermonuclear burning on the surface of the NS, giving rise to thermonuclear flashes that display a characteristic profile of a fast rise, usually to a single peak, followed by an exponential or power-law decay (see e.g. \citealt{intZand2014, Galloway2021} and references therein). X-ray bursts also repeat, with any one source showing multiple bursts during an outburst (see e.g. \citealt{Galloway2020}). After one X-ray burst, there is usually a wait time until the next event to allow time for the surface fuel layer to reform, with the exact wait time depending on the mass accretion rate. Some sources, however, show consecutive bursts with a minimal wait time (see e.g. \citealt{Keek2010}). In these cases, a fraction of the fuel layer may remain after a thermonuclear eruption and be ignited soon after the previous event. Successive bursts typically display progressively lower peaks and shorter decay times than previous bursts (see \citealt{Boirin2007} for an example), but the characteristic fast rise, exponential decay (FRED) burst profile will remain. Highly energetic Type I X-ray bursts that can produce sufficient radiation pressure to lift the optically thick surface of the NS to a larger apparent radius are known as photospheric radius expansion (PRE) bursts, and the peak of the burst typically reaches the Eddington luminosity (\citealt{Tawara1984, Lewin1984}; see also \citealt{Wolff2005} for discussion of a PRE burst from EXO 0748--676). PRE bursts generally form a small fraction of the observed Type I X-ray bursts from any one source but are valuable. Since their peak luminosity remains approximately constant at the Eddington luminosity, PRE bursts are utilised as empirical standard candles \citep{vanParadijs1978, Kuulkers2003}. 

Here, we report a sample of $22$ of Type I X-ray bursts, observed by the Rossi X-ray Timing Explorer (\textit{RXTE}), occurring \textit{during} the X-ray eclipses of EXO 0748--676 (hereafter EXO 0748). In 1985, EXO 0748 was detected in an accretion-powered X-ray outburst by the European X-ray Observatory Satellite (\textit{EXOSAT}; \citealt{Parmar1986}). EXO 0748 remained in X-ray outburst for $\sim 24$ years before entering X-ray quiescence in late 2008 (see \citealt{Degenaar2011} and references therein). During this outburst, EXO 0748 was monitored by \textit{RXTE} (\citealt{Wolff2009}; see also \citealt{Knight2023}) and \textit{XMM-Newton} (e.g. \citealt{Bonnet-Bidaud2001, Homan2003}). These observations uncovered eclipses lasting $t_e \approx 500$ s that recur on the orbital period of $P=3.824$ hrs \citep{Parmar1986, Parmar1991, Wolff2009, Knight2022a, Knight2023} and the discovery of Type I X-ray bursts which confirmed that the accretor is a NS (see \citealt{Gottwald1986, Wolff2005, Boirin2007, Paul2012} for examples). The subsequent detection of X-ray burst oscillations revealed the NS spin frequency to be within a few hertz of the measured $552$ Hz burst oscillation frequency \citep{Galloway2010}. At the time of writing, there are no known X-ray or radio pulsations at or around the burst oscillation frequency, although they are predicted \citep{Knight2023}. EXO 0748 recently returned to outburst, after $\sim 16$ years in quiescence \citep{Baglio2024, Rhodes2024} and has already exhibited several Type I X-ray bursts \citep{Kuulkers2024, Aoyama2024, Mihara2024, Bhattacharya2024, Knight2025} and eclipses \citep{Buisson2024}.

The X-ray eclipses in EXO 0748 arise as a result of the occultation of the $\sim 2 M_{\odot}$ NS \citep{Ozel2006, Knight2022a} and accretion disc by the $\sim 0.4 M_{\odot}$ M-dwarf companion \citep{Parmar1991} and its ablated outflow \citep{Knight2022a, Knight2023}. The eclipses vary in duration on long and short timescales. The ingress, egress and totality durations are all observed to vary between successive orbits, with the ingress and egress durations ranging from less than $1.0$ s to more than $30$ s \citep{Parmar1991, Wolff2009, Knight2023}. \citet{Parmar1991} suggested an X-ray-induced evaporative wind could explain the drastic variability in the ingress and egress durations, suggesting that it is necessary to sufficiently extend the eclipse transitions since the atmospheric scale height of the main sequence companion would be $\sim 100$ km and thus not produce sufficiently long eclipse transitions. Similarly, detailed studies of the X-ray eclipses exhibited by EXO 0748 strongly imply that the companion star is undergoing irradiation-driven ablation, leading to its sub-classification as a false widow. False widows are accretion-powered neutron star binaries (i.e. X-ray binaries) that ablate their companion star through X-ray irradiation \citep{Knight2023} and are named as such to highlight that the observable process of ablation is universal among false widows and spider pulsars (redbacks and black widows). The X-ray eclipses exhibited by EXO 0748 (and other false widows; see e.g. \citealt{Knight2022b}) are extended and asymmetric due to the ionised and clumpy ablated material that remains gravitationally bound to the system. These eclipse properties are analogous to the radio eclipses observed in spider pulsars. In \citet{Knight2023}, we hypothesised that false widows (mildly irradiated/ablated binary systems) represent an intermediate stage between NS LXMBs and spider pulsars (heavily irradiated/ablated binary millisecond pulsars). The premise of this hypothesis is that ablation of the companion star begins while the source is actively accreting and continues throughout the spider pulsar phase, leading to short-period binaries with very-low mass companions \citep{Knight2023}. Further support for this classification comes from the detection of a broad C IV emission line by \citet{Parikh2021}, who draw similarities between their quiescent observations of EXO 0748 and the known transitional redback pulsar, PSR J1023$+$0038, in its rotation-powered state. 

As X-ray bursts originate from the surface of the NS, they are unlikely to be directly observable during X-ray eclipse phases in sufficiently inclined systems like EXO 0748, whose inclination is $\sim 76 ^{\circ}$ (\citealt{Parmar1986}; see also \citealt{Knight2022a}), particularly as the system is shrouded by the ablated outflow. However, the collection of X-ray bursts we report on in this paper are visible \textit{during} the X-ray eclipses, raising questions regarding exactly how they are seen. In this paper, we explore the properties of the X-ray bursts seen during eclipses and across the eclipse transitions and compare them to the properties of the out-of-eclipse X-ray bursts to determine how and why these bursts are visible during eclipses and thus investigate the geometry and structure of the system. We present and categorise the different types of X-ray bursts exhibited by EXO 0748 in Section \ref{sx:Findings}. In Section \ref{sx:Stats}, we complete statistical tests on the X-ray burst population to determine how likely we are to observe these bursts from EXO 0748. In Section \ref{sx:Spectra}, we conduct a spectral analysis of these in-eclipse bursts and discuss their possible origins in Section \ref{sx:Origin}. We conclude in Section \ref{sx:Con}.

\section{Finding In-Eclipse Thermonuclear X-ray Bursts}
\label{sx:Findings}
There was extensive monitoring of EXO 0748 by \textit{RXTE} and \textit{XMM-Newton} during its initial, 24-year-long outburst (1985 - 2008). We utilise all publicly available, archival \textit{RXTE} and \textit{XMM-Newton} observations of EXO 0748 from this outburst (1985-2008). The full \textit{RXTE} data reduction procedure is described in \citep{Knight2023}, but we provide a brief overview here.

\subsection{RXTE Burst Identification}
\label{sx:Identify}

\begin{figure*}
\centering
\includegraphics[width=1\textwidth]{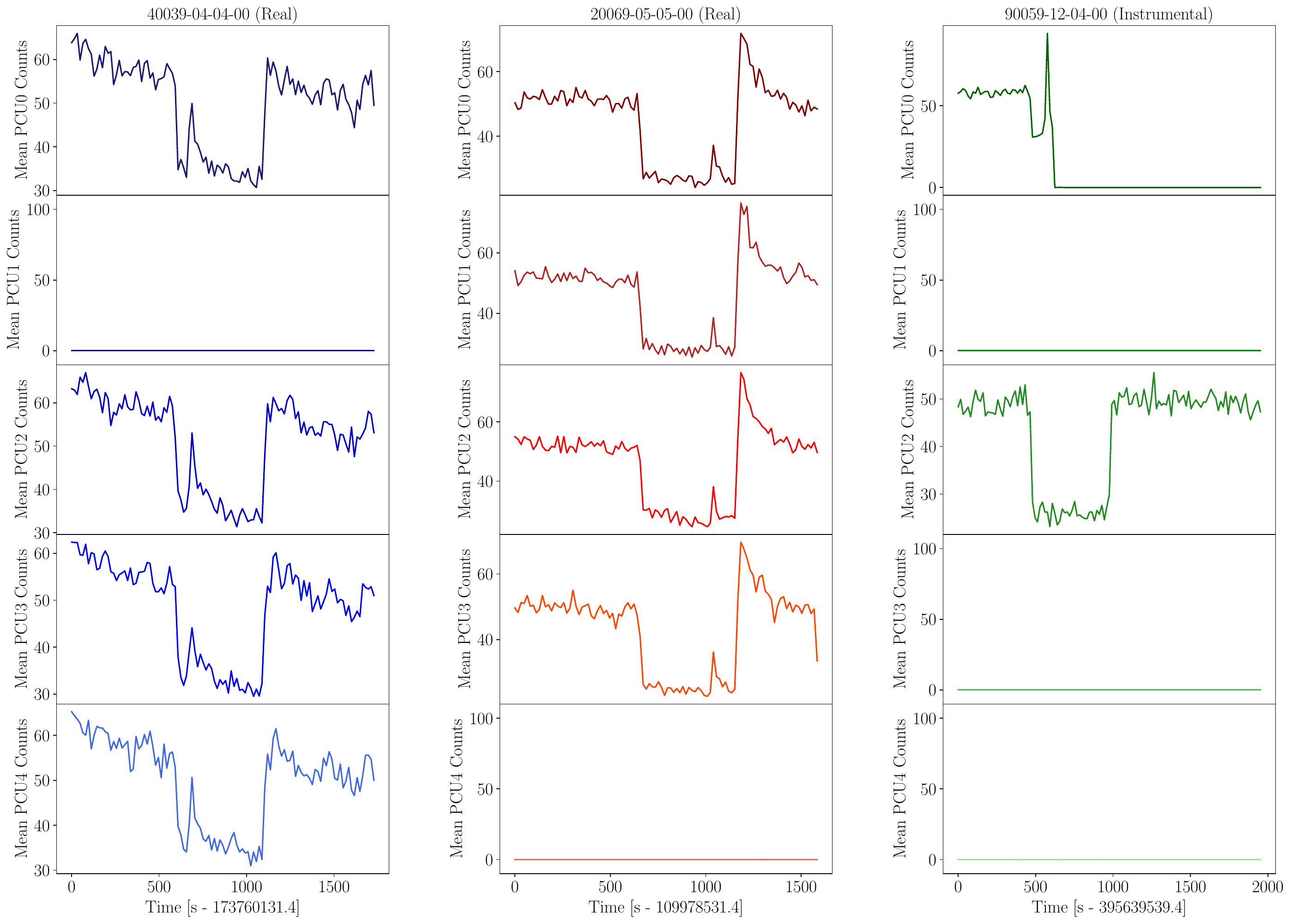}
\vspace{-0.5cm}
\caption[\textit{RXTE} Light Curves of EXO 0748--676 per PCU]{\textit{RXTE} standard-2 light curves per PCU of ObsIDs 40039-04-04-00 (left), 20069-05-05-00 (middle) and 90059-12-04-00 (right). ObsID 90059-12-04-00 demonstrates the behaviour of a malfunctioning PCU (PCU0), which creates a burst-like event before switching off and the burst-like event is not present in the other active PCU (PCU2). The in-eclipse burst in ObsID 40039-04-04-00 and the egress-split burst in ObsID 20069-05-05-00 are determined to be real as the burst is present in all active PCUs.}
\label{fig:PCU_Check}
\end{figure*}

We apply the fully automated \textsc{chromos} pipeline\footnote{\url{https://github.com/davidgardenier/chromos}} \citep{Gardenier2018} to all archival \textit{RXTE} observations of EXO 0748. The \textsc{chromos} pipeline applies all necessary data reduction steps before extracting light curves at the native time resolution of the data mode. We extract light curves in several energy bands: $3-6$ keV, $6-10$ keV, $10-16$ keV and $2-15$ keV. Each band comprises energy channels most closely matching the user-defined energy range, which accounts for the changes to the \textit{RXTE} channel-to-energy conversion throughout its lifetime. Here, we utilise the $2-15$ keV light curves, rebinned into $1$ s time bins, to identify Type I X-ray bursts. 

Since there are hundreds of archival observations of EXO 0748, we do not manually inspect each light curve. Instead, we search each time series for count rates that are $\geq 1.4$ times the average count rate in $100$ s segments of the light curve. These events are flagged and appended to an index of features (see Table \ref{tb:RXTE_Table}), which specifies the ObsID, number of active proportional counter units (PCUs), MJD at the peak of the burst and the background subtracted peak count rate. We cross-checked our index with the list of X-ray bursts from EXO 0748 presented in \citet{Galloway2008} and \citet{Galloway2020}\footnote{\url{https://burst.sci.monash.edu/}} and indicate matches in Table \ref{tb:RXTE_Table}. We inspected each event in our index to determine its nature and whether other features were flagged nearby (e.g., doublets, triplets). During this assessment, we identified several events resembling Type I X-ray bursts that coincided with or were interrupted by an X-ray eclipse. Visually, these events all displayed the characteristic fast rise -- exponential/power-law decay profile of a Type I X-ray burst, but with peak count rates generally lower than a typical Type I X-ray burst occurring during an out-of-eclipse phase. Subsequently, we search the eclipse portions of each time series with a $20 \%$ count rate threshold ($1.2$ times the average count rate) which enabled the identification of some particularly faint X-ray bursts thus enabling a full assessment of the \textit{in-eclipse} burst population.

Having identified some faint events as in-eclipse Type I X-ray bursts, it seemed prudent to determine that the identified \textit{in-eclipse} features are physical since it is unlikely that we are directly observing an X-ray burst from the surface of the NS during eclipse phases. To robustly establish which features are physical, we look at the light curves of all flagged events per PCU. Some \textit{RXTE} light curves include instrumental features that resemble X-ray bursts, which occur when a PCU breaks down\footnote{\url{https://heasarc.gsfc.nasa.gov/docs/xte/recipes/pca_breakdown.html}}. In such cases, the malfunctioning PCU exhibits a flare and then turns off (see Figure~\ref{fig:PCU_Check}, right). This behavioural pattern is unique and highly unlikely to occur in two PCUs simultaneously. Therefore, we check whether the burst pattern occurs in multiple PCUs to determine if a flagged event is physical or instrumental. Even in observations with only one active PCU, the behaviour of that PCU can indicate whether the feature is real -- if a burst-like event occurs and the PCU stays active, the burst is considered real, whereas if the PCU switches off, we classify it as a PCU breakdown. See Figure \ref{fig:PCU_Check} for an example and a list of all events excluded due to PCU breakdowns is given in Table \ref{tb:breakdowns}.

\begin{figure*}
\centering
\includegraphics[width=1\textwidth]{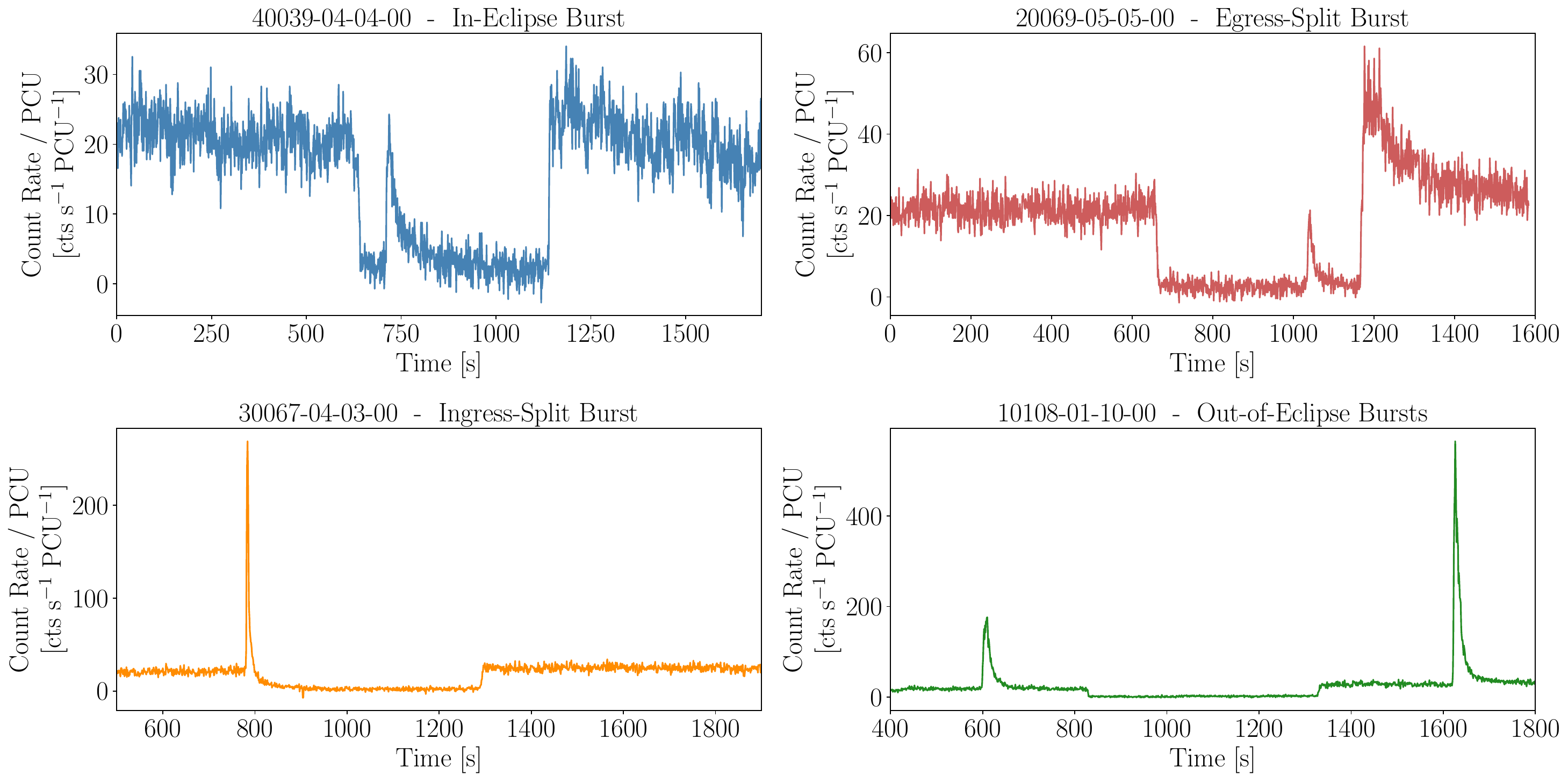}
\vspace{-0.4cm}
\caption{$2-15$ keV background subtracted \textit{RXTE} light curves depicting a representative case from each of the four groups of Type I X-ray bursts discussed in this paper. Each light curve is normalised for the number of active PCUs and the ObsID and classification are given above each panel. Figures showing all in-eclipse, egress-split and ingress-split bursts from EXO 0748--676 are provided as online supplementary material.}
\label{fig:All_Bursts_In}
\end{figure*}

After removing the instrumental flares using the PCU behaviour, we positively identify 171 thermonuclear X-ray bursts. Of these, $13$ are not listed in either \citet{Galloway2008} or \citet{Galloway2020}, and $22$ coincide with an X-ray eclipse, either fully or partially. We exclude $2$ bursts reported by \citet{Galloway2020}. These are the second burst in ObsID 92019-01-24-01, which does not display a clear fast rise -- exponential/power-law decay profile and the burst in ObsID 92019-01-25-01 as both active PCUs switched off during the burst decay phase, although we note that the peak is present in the all active PCUs. For the purposes of grouping the bursts, we define the end of burst as the time at which the average count rate first returns to pre-burst levels, using $5$s time binning. We group the population of \textit{RXTE} X-ray bursts as follows: 
\begin{enumerate}
\item In-eclipse bursts (top left, Figure \ref{fig:All_Bursts_In}), where the entire burst profile occurs within eclipse totality. We identify $9$ events in this category. 
\item Egress-split bursts (top right, Figure \ref{fig:All_Bursts_In}), where the burst starts in the later stages of totality and continues its decay phase after the egress. We identify $7$ events in this category.
\item Ingress-split bursts (lower left, Figure \ref{fig:All_Bursts_In}), where the burst starts just before or during the ingress and its decay phase is interrupted by totality. We identify $6$ events in this category.
\item Out-of-eclipse bursts (lower right, Figure \ref{fig:All_Bursts_In}), do not interrupt an eclipse at all. We identify 149 events in this category, of which $5$ are PRE bursts.
\end{enumerate}
For clarity, this classification is based solely on the orbital phase at which the burst occurs. The intrinsic properties of the bursts in all categories are assumed to be the same as they are drawn from a single underlying population of thermonuclear X-ray bursts emitted from EXO 0748. 

\subsection{XMM-Newton Burst Identification}
We reduce all \textit{XMM-Newton} EPIC observations of EXO 0748 in the XMM-Newton Science Archive (XSA) from the first outburst (all ObsIDs beginning 011, 012, 013, 016 and 021), regardless of the observing mode, to determine whether another instrument captured any in-eclipse or split X-ray bursts. The data are processed with XMM-Newton's Science Analysis Software, \textsc{xmm-sas v21.0.0} in conjunction with \textsc{heasoft v6.33} and the latest calibration files. We create EPIC-PN and EPIC-MOS event lists for each ObsID using \texttt{epproc} and \texttt{emproc}, respectively. From these event lists, we extract high-energy light curves to search for flaring events using \texttt{evselect} with the selection expression \texttt{\#XMMEA$\_$EP \&\& (PI>10000\&\&PI<12000) \&\& (PATTERN==0)} for EPIC-PN and \texttt{\#XMMEA$\_$EM \&\& (PI>10000) \&\& (PATTERN==0)} for EPIC-MOS. The EPIC-PN high-energy extraction only uses energies up to $12$ keV to avoid mistakenly identifying hot pixels as very high-energy events. Corresponding good time intervals (GTIs) for each ObsID are obtained using \texttt{tabgtigen} with the rate expression determined by identifying a count rate threshold just above the mean level in the background light curves, thus differing for each observation. We apply the GTI filters using \texttt{evselect} with the selection expressions \texttt{\#XMMEA$\_$EP \&\& gti(obsid$\_$gti.fits,TIME) \&\& (PI>150)} and \texttt{\#XMMEA$\_$EM \&\& gti(obsid$\_$gti.fits,TIME) \&\& (PI>150)} respectively for EPIC-PN and EPIC-MOS. Here \texttt{obsid$\_$gti.fits} is the GTI filter file produced by \texttt{tabgtigen} for each ObsID. Subsequently, we extract source light curves in the $0.5 - 10.0$ keV range using circular source regions manually determined using \textsc{ds9} for any data taken in imaging mode. The same approach determines an appropriate rectangular source region for timing mode observations. The time binning used for each light curve depended on the instrument mode, and so varied for each ObsID. We search the light curves for X-ray bursts using the procedure described in Section \ref{sx:Identify} using a $20 \%$ count rate threshold. Within the \textit{XMM-Newton} archival data, we identify $112$ out-of-eclipse bursts, including doublets and triplets, and $0$ in-eclipse or split bursts. As listed in Table \ref{tb:RXTE_Table}, there are $3$ cases where \textit{XMM-Newton} captured the same out-of-eclipse bursts as \textit{RXTE}.

\begin{figure*}
\centering
\includegraphics[width=0.97\textwidth]{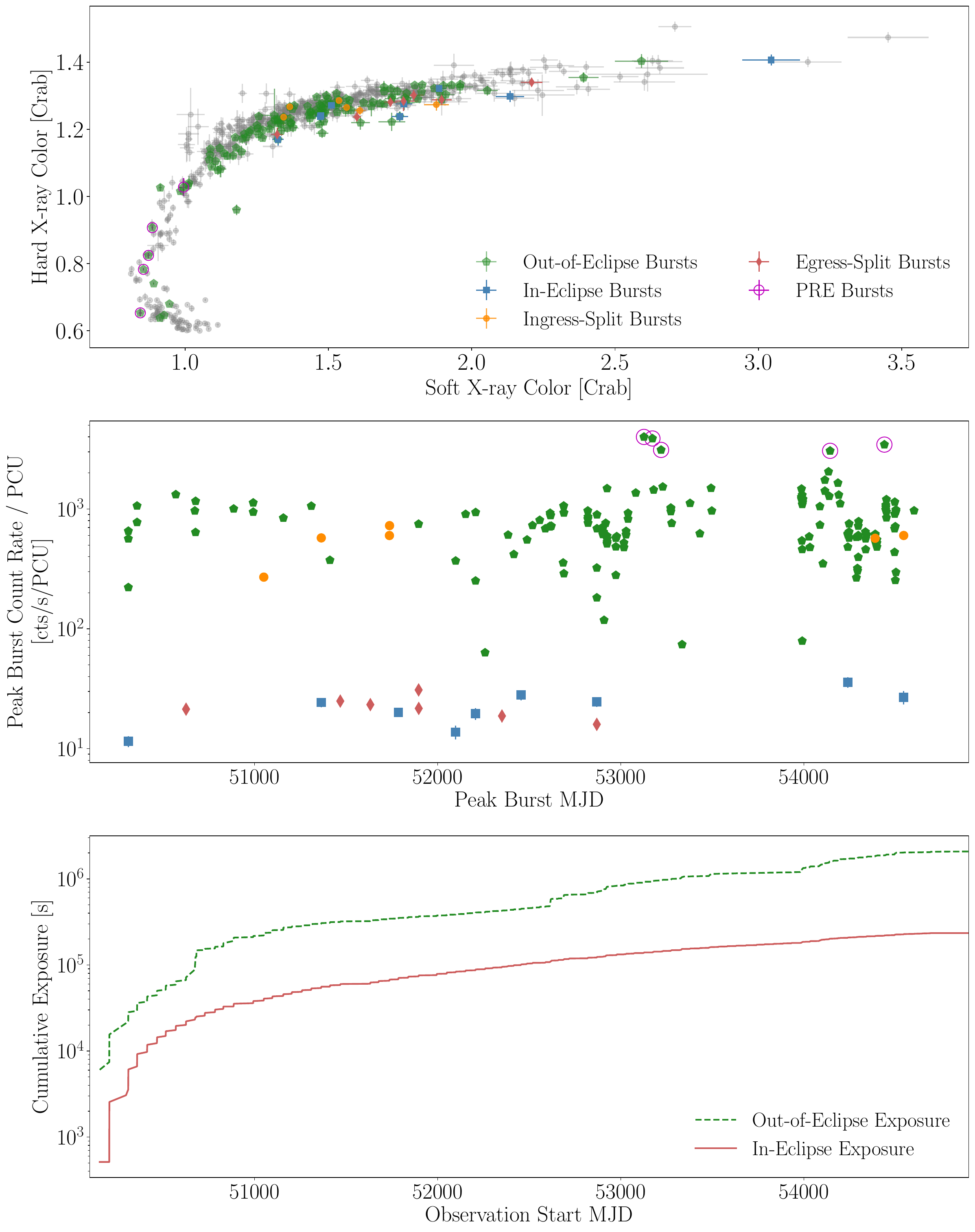}
\vspace{-0.15cm}
\caption{Top: A colour-colour diagram of all \textit{RXTE} observations of EXO 0748--676, showing the majority of observations form the extreme island (hard state), while a smaller fraction form the banana state (soft, lower left). Highlighted on the diagram are the different groups of X-ray bursts discussed in this paper. Middle: A distribution of the background subtracted, peak count rate per active PCU, in the $2 - 15$ keV band, for the all the Type I X-ray bursts observed from EXO 0748--676 by \textit{RXTE}. Bottom: The cumulative in-eclipse (solid line) and out-of-eclipse (dashed) \textit{RXTE} exposure of EXO 0748--676.}
\label{fig:ccd}
\end{figure*}

\section{Burst Statistics}
\label{sx:Stats}
We can deduce the nature of the in-eclipse bursts by analysing the number of in-eclipse (ec) and out-of-eclipse (ooe) bursts detected and their relative peak count rates. Here, we complete several statistical tests to further understand these events within the \textit{RXTE} and \textit{XMM-Newton} burst populations.

For \textit{RXTE}, we detect $N_{\rm{ec, peak}} = 16$ bursts that peak during totality (of which $N_{\rm{eg}} = 7$ are egress-split burst), and $N_{\rm{ooe, peak}}=155$ that peak out-of-eclipse (of which $N_{\rm{in}} = 6$ are ingress-split bursts). We show this population of bursts in Figures \ref{fig:All_Bursts_In} - \ref{fig:ccd} inclusive and provide a full index of bursts in Table \ref{tb:RXTE_Table}. We further identify $5$ of the out-of-eclipse bursts as PRE bursts. Many of these X-ray bursts appear in the database of thermonuclear bursts by \citet{Galloway2008, Galloway2020}, and some of the X-ray bursts that coincide with an eclipse appear in the \citet{Wolff2009} database of eclipse timings. If an event appears in either of these databases, we note it in Table~\ref{tb:RXTE_Table}. Events listed in Table \ref{tb:RXTE_Table} without citation do not appear in either database but may have been studied independently. For \textit{XMM-Newton}, we detect no bursts during eclipses and $N_{\rm{ooe, peak}}=112$ out-of-eclipse bursts. 

\subsection{The Detection Fraction of In-Eclipse Bursts}
Here we compare the \textit{observed} number of bursts peaking during totality to the number of bursts \textit{expected} to peak during totality, given the number of out-of-eclipse bursts and the time spent observing in and out of eclipse ($T_{\rm{ec}}$ and $T_{\rm{ooe}}$ respectively). This expectation value is given as $N_{\rm{expected}} = (T_{\rm{ec}} / T_{\rm{ooe}}) N_{\rm{ooe, peak}}$, where $N_{\rm{ooe, peak}}=155$ for \textit{RXTE} and $N_{\rm{ooe, peak}}=112$ for \textit{XMM-Newton}, as determined above. We note that computing the expectation value in this way assumes a constant burst rate throughout. As such, we assess the validity of this assumption by determining the spectral state of the source at the time of each burst, since the burst rate will depend on the accretion rate and thus the spectral state. We determine that all in-eclipse (blue squares), egress-split (red diamonds) and ingress-split (orange circles) X-ray bursts occurred while EXO 0748 was in the hard spectral state (island state / extreme island state\footnote{Atoll sources exhibit a hard, power-law spectra at low luminosities when in the island state. In the extreme island state, sources typically show a decrease in soft colour intensity while the hard colour intensity is stable.}; hard X-ray colours $\gtrsim 1.0$; e.g \citealt{Mancuso2019}), as shown in the top panel of Figure \ref{fig:ccd}. Here, the soft and hard X-ray colours are, respectively, calculated as the ratio of counts in the energy bands $3.5 - 6.0 / 2.0 - 3.5$ keV and $9.7-16.0/ 6.0-9.7$ keV, with the bursts, eclipses and dips removed. We also see from Figure \ref{fig:ccd} that that majority of the \textit{RXTE} observations of EXO 0748 and $\sim 95 \%$ of the detected bursts occurred during the hard spectral state. Therefore our assumption of a constant burst rate is sufficient here.

To determine $T_{\rm{ec}}$ and $T_{\rm{ooe}}$ we utilise our previously published, simple eclipse model \citep{Knight2023}, which fits a series of straight lines between the four eclipse contacts (see also \citealt{Wolff2009}). In this model, the eclipse contacts are the start of the ingress, $t_{1}$, the start of totality, $t_{2}$, the end of totality, $t_{3}$ and the end of the egress, $t_{4}$. As such, the in-eclipse exposure time is $t_{3} - t_{2}$, the ingress duration is $t_{2} - t_{1}$ and the egress duration is $t_{4} - t_{3}$ \citep{Knight2023}. The eclipse contacts were manually adjusted to the last available time bin if the ObsID contained a partial eclipse. The out-of-eclipse exposure time is thus the total good exposure time minus the in-eclipse exposure. Note that using the ratio of in-eclipse to out-of-eclipse exposure time corrects for any observational bias arising from the large number of observations of EXO 0748 that targeted the eclipses \citep{Wolff2009}. To the nearest second, we find $T_{\rm{ec}} = 234374$ s and $T_{\rm{ooe}} = 2086644$ s for \textit{RXTE} (in-eclipse exposure is $\sim 10 \%$ of total exposure), and $T_{\rm{ec}} = 33323$~s and $T_{\rm{ooe}} = 999043$ s for \textit{XMM-Newton} (in-eclipse exposure is $\sim 3 \%$ of total exposure).

The resulting expected number of in-eclipse bursts is $N_{\rm{expected}} = 17.41$ for \textit{RXTE}, and $N_{\rm{expected}} =3.74$ for \textit{XMM-Newton}. Both of these values are larger than the observed number of in-eclipse bursts: $N_{\rm{ec, peak}} = 16$ for \textit{RXTE} and $N_{\rm{ec, peak}} = 0$ for \textit{XMM-Newton}. This discrepancy could be purely down to Poisson statistics, or it could be because we fail to detect some bursts that occur during totality whilst we are observing. The missed bursts are presumably bursts that are too faint to distinguish from the residual in-eclipse flux.

\begin{figure}
    \centering
    \includegraphics[width=\columnwidth]{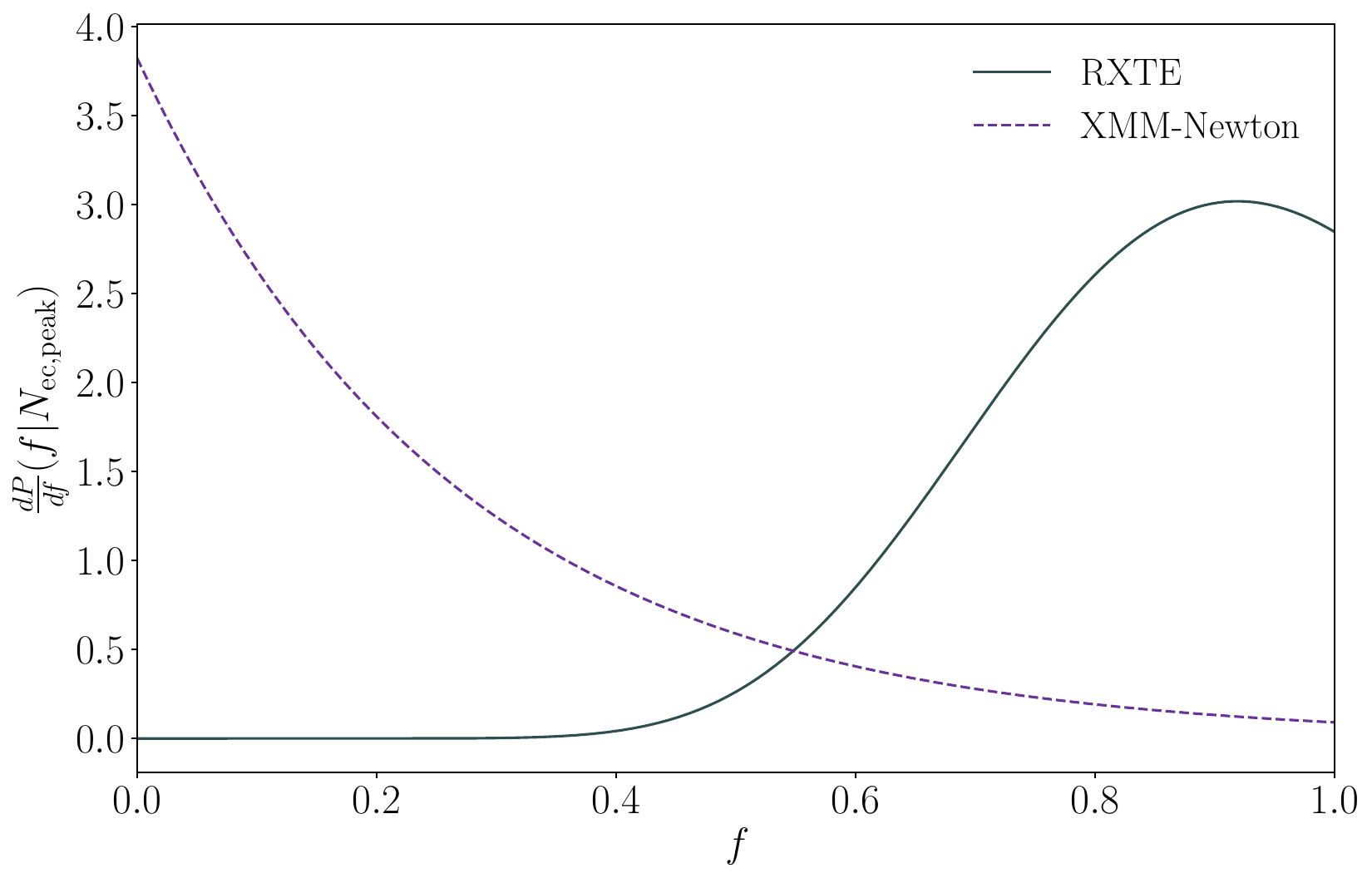}
    \vspace*{-0.5cm}
    \caption[]{The probability density of the fraction $f$ given that we detect $N_{\rm ec,peak}=16$ X-ray bursts that peak during eclipse with \textit{RXTE} (grey, solid) and $N_{\rm ec,peak}=0$ with \textit{XMM-Newton} (purple, dashed). Here, $f$ is the fraction of bursts occurring during an observed eclipse that we are able to detect.} 
    \label{fig:p_f_k}
\end{figure}

If we are only able to detect a fraction, $f$, of the bursts that occur during observed eclipses, then the number of bursts we expect to detect during eclipse is $f N_{\rm expected}$, and the probability of detecting $N_{\rm ec,peak}$ bursts during totality is
\begin{equation}
    P(N_{\rm ec,peak}|f) = \exp(- f N_{\rm{expected}}) \frac{(f N_{\rm expected})^{N_{\rm ec,peak}}}{N_{\rm ec,peak}!}.
\end{equation}
This is simply the Poisson likelihood of $N_{\rm ec,peak}$ for a given $f$, and we can use it to place statistical limits on the fraction $f$ for both \textit{RXTE} and \textit{XMM-Newton}. To do so, we use Bayes' theorem to derive the probability density of $f$ given $N_{\rm ec,peak}$ (the \textit{posterior}). Assuming a flat prior on $f$, the posterior simply becomes
\begin{equation}
    \frac{dP}{df}(f|N_{\rm ec,peak}) \propto P(N_{\rm ec,peak}|f),
\end{equation}
where the constant of proportionality is set to ensure that the integral of the posterior from $f=0$ to $f=1$ is unity. Figure \ref{fig:p_f_k} shows the resulting posteriors for \textit{RXTE} (black, solid) and \textit{XMM-Newton} (purple, dashed).

We derive the desired limits on $f$ by integrating these distributions from the peak until we reach a target confidence level. For $1\sigma$ confidence ($68.27\%$), we find $f =0.92^{+0.082}_{-0.158}$ for \textit{RXTE} and $f \leq 0.29$ for \textit{XMM-Newton}. For $90\%$ confidence, these numbers become $f=0.92^{+0.082}_{-0.284}$ and $f \leq 0.56$. We therefore see that the fraction $f$ is larger for \textit{RXTE} than for \textit{XMM-Newton} with $>90\%$ statistical confidence. We suggest that this due to the softer band pass of \textit{XMM-Newton} being more heavily affected by absorption, but is also likely affected by the observing mode and sensitivity of each EPIC detector.

Finally, we note that our use of Poisson statistics within this section, while reasonable, cannot be strictly correct because the eclipse interval is comparable to the recurrence  time of the X-ray bursts and the bursts are not occurring independently. Therefore, X-ray bursts likely have a quasi-periodicity that could sometimes beat with the eclipse periodicity. Exploring this phase dependence, however, is complex and beyond the scope of this paper.

\subsection{Flux Distributions}
To further understand the population of \textit{RXTE} X-ray bursts, we consider Figure \ref{fig:ccd}, which shows the background subtracted peak burst count rate per active PCU against the MJD of the burst peak for the different groups of X-ray bursts discussed in this paper. We uncover a dichotomy between the in-eclipse (blue squares) and egress split (red diamonds) bursts and the out-of-eclipse (green pentagons) and ingress split (orange circles) bursts. The former two classes have burst peaks during the totality, so their peak count rates are relatively low. In contrast, the latter two classes peak during out-of-eclipse phases, so the peak count rates are comparatively high. We note that the low peak count rates exhibited by some of the out-of-eclipse bursts are due to either consecutive events, for example, doublets and triplets, or observations that captured the burst shortly after the peak. We further note that our reported peaks for several egress-split bursts differ from those in \citet{Galloway2020} as our peak is determined by the in-eclipse portion of these bursts.

Despite consistent monitoring of EXO 0748 during the \textit{RXTE} era (see Figure \ref{fig:ccd}, bottom panel), we note a lack of observed X-ray bursts between $53500 - 54000$ MJD. We do not identify any in-eclipse, ingress-split or egress-split bursts during this time, and only a small number of out-of-eclipse bursts. One likely explanation for the lower number of detections in this interval is the change in spectral state known to occur around this time (see e.g. \citealt{Ponti2014_specstate, Knight2023}), which coincides with an increase in the observed flux. Assuming the increased flux is a consequence of an increased mass accretion rate, the accreted fuel layer could have undergone periods of stable burning, thus increasing the wait times between bursts (see \citealt{Boirin2007} and references therein) and making detections of in-eclipse bursts less likely.

Alternatively, the large gap in detection of in-eclipse bursts may arise from the quasi-periodicity of the X-ray bursts i.e. the bursts and eclipses could be partially phase-locked between $53500 - 54000$ MJD, inhibiting the detection of X-ray bursts. Another possibility is that, during this period, a smaller fraction of the burst flux was being scattered around the companion star into our line of sight, due to changes in the nature of the scattering material. In this context, it is interesting to note that the lack of observed X-ray bursts between $53500 - 54000$ MJD occurs when the eclipse asymmetry reversed (when the ingress was longer than the egress; see \citealt{Knight2023}). The reversal of the eclipse asymmetry is thought to be driven by the movement of the gravitationally bound ablated material in the system, which would influence the scattered fraction if the ablated material plays the role of the scattering medium.

\begin{figure}
    \centering
    \includegraphics[width=\columnwidth]{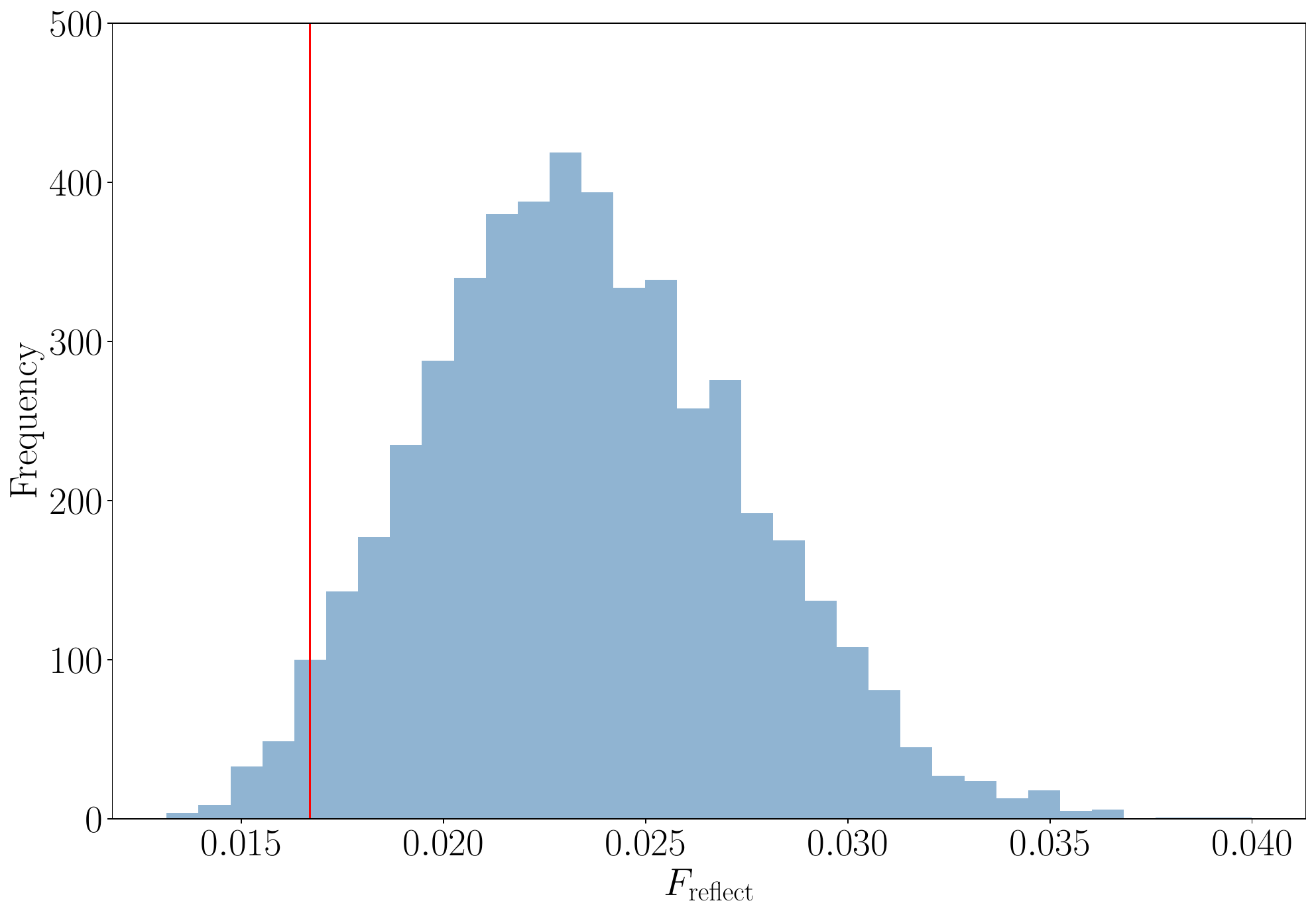}
    \vspace*{-0.5cm}
    \caption{A histogram depicting $5000$ calculations of the ratio of the mean peak count rate of the X-ray bursts that peak during totality (in-eclipse and egress-split bursts) to the mean peak count rate of the X-ray bursts that peak out-of-eclipse using a random sample of $16$ bursts that peak during out-of-eclipse phases. The mean reflection fraction of the bursts is $F_{\rm{reflect}} = 0.024$ and the corresponding standard deviation is $\sigma = 0.004$. The red line depicts the quiescent reflection fraction of $0.0167$.}
    \label{fig:refhist}
\end{figure}

The likely reason for us being able to observe bursts during totality is that some fraction of the burst emission from the neutron star surface is reflected into our line of sight by a large-scale scattering medium surrounding the system (with the rest either passing straight through or being absorbed and re-emitted at longer wavelengths; \citealt{Paul2012, Hynes_multiwavelength_2006, Knight2025}). For bursts observed out of eclipse, we see both the direct and reflected emission, whereas during eclipse, we only see the reflected emission. We can use the observed population of bursts from Fig. \ref{fig:ccd} to constrain the \textit{reflection fraction}, $F_{\rm reflect}$, which is the fraction of the total flux we observe from an out-of-eclipse burst that is reflected by the scattering medium.

Under this definition, $F_{\rm reflect}$ is equal to the peak count rate observed from an in-eclipse burst divided by the peak count rate that the same burst \textit{would have had} if it had instead been observed out-of-eclipse. The former can be estimated from the mean peak count rate of the 16 bursts observed to peak during totality, $C_{\rm ec,peak}$ (all background subtracted and expressed per active PCU). The latter can be estimated from the peak count rates of the 155 bursts observed to peak out of eclipse (again, background subtracted and per PCU), but we also need to account for the fact that some of these bursts were not bright enough to have been detected had they occurred during totality. To do this, we could calculate the mean peak count rate of the brightest $f \times 155$ bursts, where we measured $f =0.92^{+0.082}_{-0.158}$ in the previous subsection. However, this would not account for the \textit{distribution} of burst peak count rates, which is non-Gaussian due to e.g. a few very bright but very rare PRE bursts and some dimmer bursts that occur in doublets and triplets. We therefore account for the true distribution of burst count rates, and the derived posterior probability distribution of $f$, by running a Monte Carlo simulation. For each step in the simulation, we
\begin{enumerate}
\item randomly draw a value for $f$ from its derived probability distribution (Fig. \ref{fig:p_f_k});
\item randomly select 16 peak count rates from the $f \times 155$ brightest bursts observed to peak out of eclipse;
\item calculate the mean peak count rate of these 16 randomly selected bursts $C_{\rm select}$; and
\item calculate a reflection fraction estimate as $F_{\rm{reflect}} = C_{\rm ec, peak} / C_{\rm select}$.
\end{enumerate}
We run the simulation for 5000 steps, yielding 5000 values for $F_{\rm reflect}$ that we plot in the histogram shown in Fig. \ref{fig:refhist}. This calculation yields a mean reflection fraction of $F_{\rm reflect}= 0.024 \pm 0.004$ ($1 \sigma$ uncertainty), such that $\approx 2.4\%$ of the flux from a typical burst is reflected. In other words, we estimate that the 16 bursts observed to peak during totality would have been a factor $1/F_{\rm reflect}\approx 42$ brighter had they occurred out of eclipse.

For comparison, we compute the reflection fraction of the persistent emission as the ratio of the average background subtracted, in-eclipse count rate per PCU to the average background subtracted out-of-eclipse count rate per PCU as, $R = C_{\rm{tot}} / C_{\rm{ooe}} = 0.0167$, using over $400$ eclipses \citep{Wolff2009, Knight2023}. This level is represented by the red line in Figure \ref{fig:refhist} showing that the reflection fraction of the bursts typically exceeds that of the persistent emission, although the two are consistent within $2~\sigma$ confidence. Note that all count rates used here are background subtracted rates in the $2-15$ keV band, per active PCU, thus justifying the difference between this and the residual in-eclipse flux levels previously reported. For example, \citet{Parmar1986}, report a residual flux level $\sim 4\%$ in the $2-6$ keV band from early EXOSAT observations, which they interpret as originating from two components; one contributing at energies $\leq 2$ keV and having a very soft spectrum, while the other at energies over $2$ keV with a spectrum similar to that of the quiescent emission. Thus, we deduce that the soft X-ray contribution to the residual eclipse emission is negligible for \textit{RXTE} data while for \textit{XMM-Newton's} EPIC detectors, which probe softer X-ray emission, the residual emission in-eclipse is typically higher.

Our analysis shows that $F_{\rm reflect} \approx 2.4 \%$ is larger than $R \approx 1.67 \%$, suggesting that the scattering of the X-ray bursts peaking during totality is special, such that more flux scatters during a burst than otherwise. The probability that $R \leq 0.0167$, given that R belongs to the distribution in Figure \ref{fig:refhist} is $p = 0.0028$, indicating that it is unlikely that an in-eclipse X-ray burst will be observable with a reflection fraction less than $R$. This, suggests that we are not simply observing bright X-ray bursts during eclipses and we actually observe a greater fraction of the burst flux during an eclipse than the non-burst flux. Therefore, extra scattering of the burst emission into our line of sight appears necessary to explain the fraction of the in-eclipse bursts we observe. This behaviour may also be explained if the efficiency of the reflection is energy dependent such that there are more hard photons than soft photons during the bursts. Since harder X-rays reflect more easily than soft photons, this could lead to higher reflection fractions during the bursts. Another explanation is that there is an increase in the quantity and/or extent of the scatterer during the X-ray bursts. Analysis by \citet{Keek2016} determined that the reflection fractions are larger in scenarios whereby the inner accretion disc increases steeply in height, causing some of the companion star to be obscured. The authors note that such a geometry could be induced by the X-ray burst itself, if X-ray heating causes the inner disc to puff up. This scenario is consistent with our findings, although exploring this fully is beyond the scope of this paper.

\section{Spectral Analysis}
\label{sx:Spectra}

\begin{figure*}
    \centering
    \includegraphics[width=\textwidth]{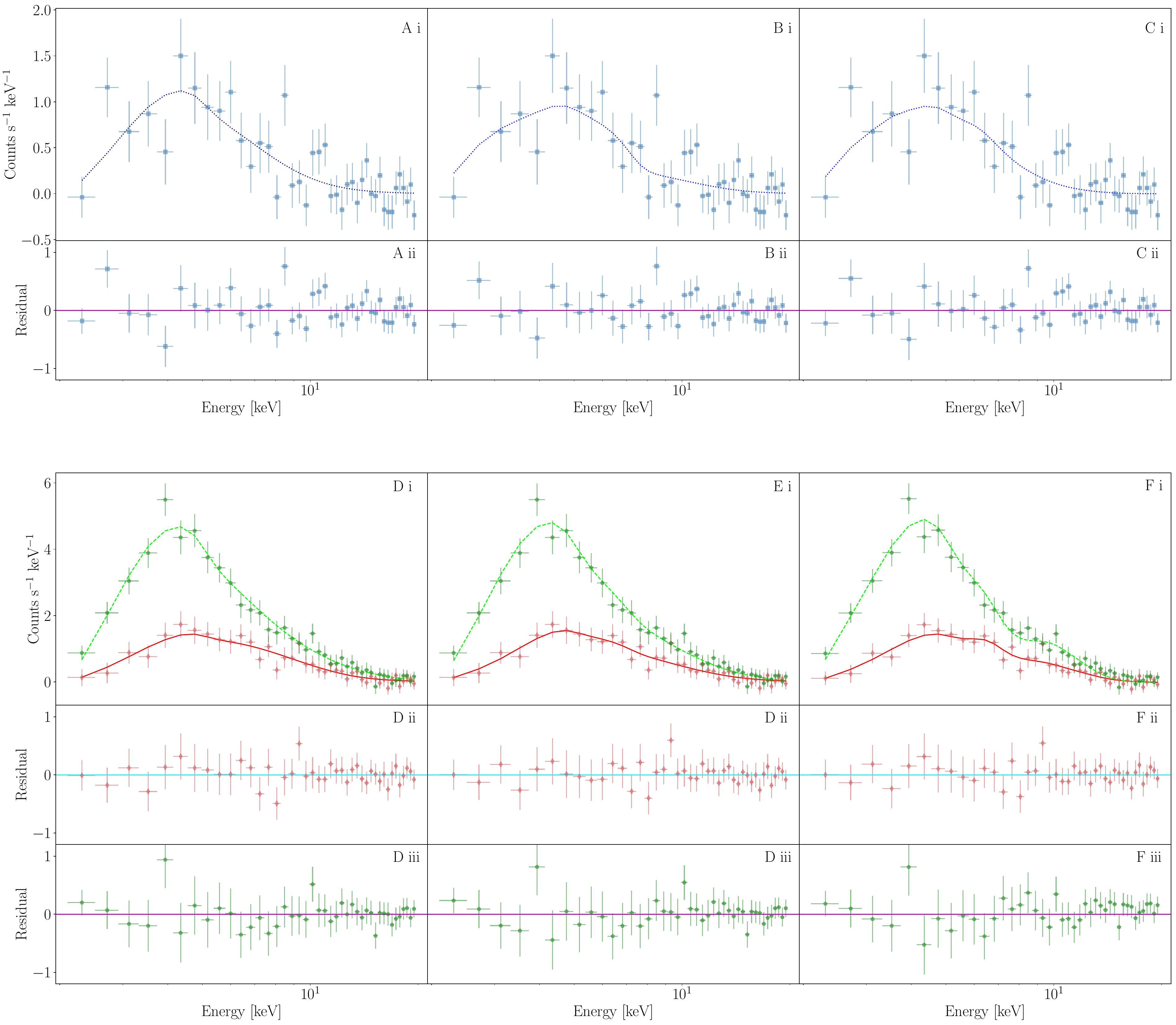}
    \vspace*{-0.5cm}
    \caption{Epoch 4 \textit{RXTE} PCU2 spectra, extracted from $64$s periods surrounding the peak of the in-eclipse (A, B, and C) and egress-split X-ray bursts (diamonds in D, E and F) with 3 different models. The models are: 1) \texttt{tbabs*(diskbb+bbodyrad)} (A, D) 2) \texttt{tbabs(zxipcf*pexriv)} (B, E) and 3) \texttt{tbabs(zxipcf*xillverns)} (C, F). The out-of-eclipse tails of egress-split bursts (pentagons in D, E and F) are fit with the same three models. Residuals from each fit are show in the lower panels (ii and iii). Here, we demonstrate that the signal-to-noise prevents statistical distinction between the models.  
    }
    \label{fig:in-spec}
\end{figure*}

In this section, we perform spectral fits to the X-ray bursts observed to peak during totality, to determine whether they require a reflection component, thereby supporting the scenario whereby the burst emission is reflected into our line-of-sight during eclipses.

For all \textit{RXTE} X-ray bursts observed to peak during totality, we extract $2 - 25$ keV PCU2 spectra from a $64$ s time range centered on the peak of the burst. For the egress-split bursts, we also extract $2 - 25$ keV PCU2 spectra of the burst decay from a $96$ s time range starting from the first time bin after egress. For each ObsID we create a filter file using \texttt{xtefilt} and a GTI file from the user-specified, time range using \texttt{timetrans}. We extract standard-2 source spectra using \texttt{saextract} with $16$ s time bins, correct for deadtime and create appropriate response matrices using \texttt{pcarsp}. We follow the same procedure to extract spectra from $100$ s sections of totality that do not contain a burst. These spectra serve as the background spectra for the in-eclipse burst, as they will represent the background at the time of an in-eclipse X-ray burst more accurately than a typical standard-2 background spectra. For the spectra of the burst decay, we utilise select $100$s portions of the out-of-eclipse light curves to serve as background spectra to ensure all spectra have the persistent emission subtracted. 

For \textit{RXTE}, the energy-channel conversions changed during the course of the mission due to gain changes for the Proportional Counter Array (PCA) and the loss of propane for PCU0, thus preventing simultaneous fits to all spectra. Instead we perform simultaneous spectral fitting within each gain epoch using \textsc{xspec} version 12.13.1 \citep{Arnaud1996}. For the in-eclipse bursts we group the bursts as follows: 10108-01-06-00 (Epoch 3), 40039-04-04-00 (Epoch 4), and all other in-eclipse ObsIDs (Epoch 5). Similarly for the egress-split bursts: 20069-05-05-00 (Epoch 3), 40039-06-01-00 and 50045-01-04-00 (Epoch 4) and 50045-06-05-00 and 70048-13-06-00 (Epoch 5). We do not utilise ObsIDs 50045-06-02-00 and 70048-02-04-00 here as the $16$~s standard-2 time resolution prevents clear detection of the peak of these egress-split bursts, which in both cases occurs very close to or during the egress.

For all groups we trial three different \textsc{xspec} models : 1) \texttt{tbabs*(diskbb + bbodyrad)} (Figure \ref{fig:in-spec}, panels A and D); 2) \texttt{tbabs(zxipcf*pexriv)} (Figure \ref{fig:in-spec}, panels B and E); and 3) \texttt{tbabs(zxipcf*xillverns)} (Figure \ref{fig:in-spec}, panels C and F). In these models, \texttt{tbabs} calculates the cross section for X-ray absorption by the ISM which we fix to $0.149 \times 10^{22}$ for all fits (e.g. \citealt{Knight2022a}). \texttt{diskbb} is a multi-temperature accretion disc blackbody spectrum \citep{Makishima1986}, and \texttt{bbodyrad} is a NS surface blackbody spectrum with the normalisation proportional to the surface area of the emitter. \texttt{zxipcf} considers absorption by an ionised material with partial covering \citep{Miller2006}, which we have previously utilised to model the ablated outflow in EXO 0748 (see \citealt{Knight2022a} and \citealt{Knight2023}). \texttt{pexriv} \citep{Magdziarz1995, Done1992} and \texttt{xillverns} are both models for reflection spectra. The former is a reflected power law for an ionised material, while the latter is a more sophisticated model for reflection of an incident blackbody spectrum, and also considers fluorescence lines \citep{Garcia2014, Dauser2014, Dauser2022}. For both reflection models, we fix the inclination angle to the known value of $76~^{\circ}$ and provide all best fitting model parameters in Table \ref{tb:spec}.

As demonstrated in Figure \ref{fig:in-spec}, which shows the resulting fits to the epoch 4 bursts with each of the three models, the signal-to-noise of the extracted spectra does not allow the models to be separated statistically. For example, the epoch 5 group of in-eclipse bursts achieved $\chi/\nu = 1.11 $, $\chi/\nu = 0.99$ and $\chi/\nu = 1.03$, respectively, for models 1, 2 and 3 described above. The corresponding null hypothesis probabilities are $0.166$, $0.156$ and $0.372$. The same is true for the spectral fits to the peaks of the egress-split bursts in epoch 5, where we obtain $\chi/\nu = 1.04$, $\chi/\nu = 1.05$ and $\chi/\nu = 1.07$, respectively, for models 1, 2 and 3, and corresponding null hypothesis probabilities are $0.293$, $0.135$ and $0.350$. As demonstrated in Figure \ref{fig:in-spec}, this inability to distinguish between models is also true for the fits in other gain epochs as the different models all show similar spectral residuals. However, for the spectral fits to the out-of-eclipse decay tails of the egress-split bursts (green pentagons in Figure \ref{fig:in-spec}), we can favour model 1 (non-reflection model) for the epoch 5 bursts which achieves $\chi/\nu = 1.22$,  $\chi/\nu = 1.45$ and $\chi/\nu = 3.52$, respectively, for models 1, 2 and 3. However, this is not the case for the fits to the burst tails in epochs 3 or 4 (see Table \ref{tb:spec} for details).

Despite the signal-to-noise of the bursts that peak during totality preventing a confident distinction between the three trialed models, there are a few points of interest. Firstly, all fits using the non-reflection model (model 1) feature a weak or negligible contribution from an accretion disc (the strength of \textsc{diskbb} components is close to or consistent with zero), which is not surprising for burst spectra that are dominated by the blackbody flash. However, even the fits to the out-of-eclipse decay tails of the egress-split bursts, which occur after the eclipse so we would expect the disc to reemerge, do not require strong disc contributions. This suggests that even if the accretion disc physically extended beyond the companion star, the disc emission is not reaching the observer, or the emission from the outer disc is too cool to be visible in an X-ray spectrum. This supports earlier analysis by \citet{Knight2022a, Knight2023} which suggested that ablated material surrounds the companion star in EXO 0748--676 and blocks our view of the accretion disc for a short time either side of the eclipse. When comparing the best-fitting parameters of both reflection models (middle and lower panels of Table \ref{tb:spec}), we find that the measured column density of the absorber is typically $N_{\rm{H}} > 10^{23}$ cm$^{-2}$, the covering fraction of the absorber is close to unity in most cases, and the ionisation of the absorber modeled with \texttt{zxipcf} is typically $\log(\xi) > 2.0$. These properties are compatible with the ablated material \citep{Knight2023} and are equally justified by an accretion disc wind \citep{Tomaru2023}, so either could also be a plausible reflection site for the X-ray bursts peaking during totality. Although, we note that higher values of $\log(\xi)$ are often associated with disc winds \citep{Datta2024}.

\begin{figure*}
\centering
\includegraphics[width=\linewidth]{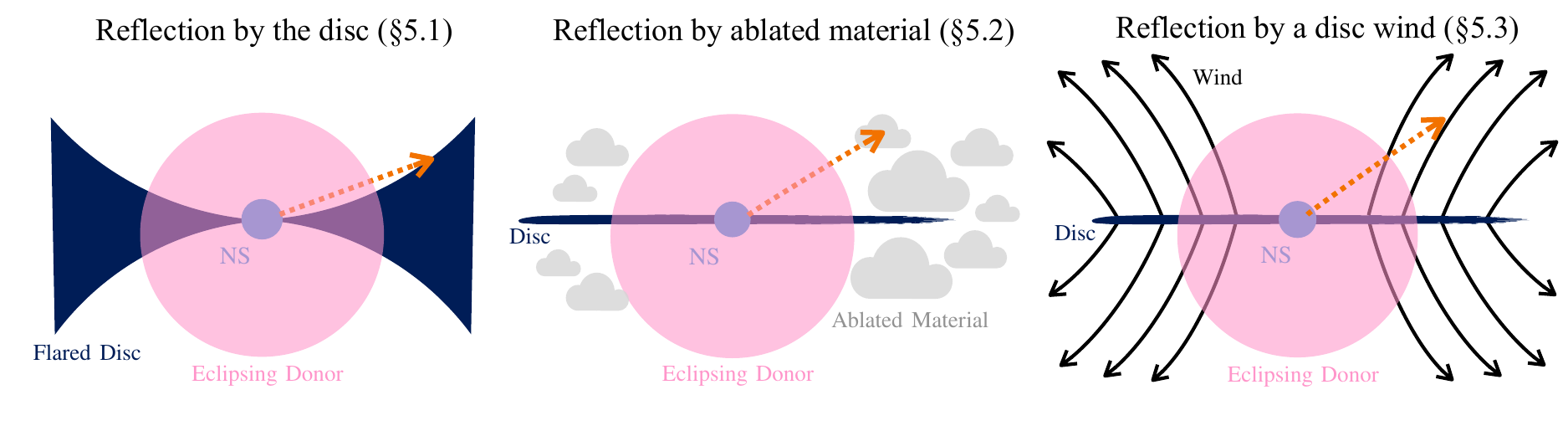}
\caption{Schematic diagram showing the three scenarios we consider (in Section \ref{sx:Origin}) as possible origins of the in-eclipse bursts. In each case, the schematic gives a very approximate, not to scale observer viewpoint during the eclipse (zero orbital phase). The orange dotted arrow shows the incident burst emission interacting with the reflection site, which is an un-eclipsed structure in each case. Any reflected emission is thus directed out of the page and towards the observer.}
\label{fig:scenarios}
\end{figure*}

\vspace*{-2cm}
\section{Possible Reflection Sites of the In-Eclipse X-ray Bursts}
\label{sx:Origin}

We consider three scenarios, illustrated in Fig. \ref{fig:scenarios}, for explaining how we observe X-ray bursts during eclipses and across the eclipse transitions. We discuss each in turn.

\subsection{Reflected by the Accretion Disc}
To assess whether the in-eclipse X-ray bursts in EXO 0748 could arise due to reflection off of the outer accretion disc, we begin by calculating how much of the disc is visible at different orbital phases. Let us assume a spherical companion star and the prescription of \citet{Knight2022a} to compute the companion star's radius. We approximate the accretion disc as tidally truncated at $\sim 0.9$ its own Roche Lobe radius, e.g., $a-r_{\rm cs}$, such that the outer disc radius is $r_{\rm out} = 0.9 (a-r_{\rm cs})$ \citep{Mushtukov2019}. We assume a mass ratio of $q=0.222$ and viewer inclination angle $i=76.5^\circ$, as is appropriate for EXO 0748 \citep{Knight2022a}. From this, we find that even at the centre of totality, the outer edges of the accretion disc are visible while the X-ray bright, central portion of the disc is obscured. Thus, some fraction of the disc emission still reaches the observer.

The fraction of the emission reaching the observer will be contributed to by X-rays originating from the NS and inner disc that then reflect off the outer disc. To explore this contribution and assess whether it is a possible origin for the bursts peaking during totality, we compute the resulting reflection fraction and compare it to the value calculated from the observations in Section \ref{sx:Stats}. To do this, we assume a simple model that ignores relativistic effects and assumes a point-like, isotropic `lamp post' source located at a height $h$ above the NS as the illuminator. For a flat accretion disc, the reflected flux we observe per unit disc radius as a fraction of the direct flux we see is given by
\begin{equation}
    \frac{dF}{dr} = 2 \cos i \frac{h r}{(h^2+r^2)^{3/2}}.
    \label{eqn:dfdr_flat}
\end{equation}
This assumes that all flux incident on the disc is re-emitted isotropically and we do not account for the anisotropy of the direct source or of the reflector. While the above is an expression for the bolometric flux, we make the simplifying approximation that we can use it to describe the flux in the \textit{RXTE} bandpass.

We set $h=r_{\rm NS}$, where $r_{\rm NS}$ is the radius of the NS, to represent the bulk of the emission originating from the NS surface. Integrating the flux from an inner radius of $r_{\rm NS}$, to an outer radius of infinity gives
\begin{equation}
    F = \int_{r_{\rm{NS}}}^\infty \frac{dF}{dr}~dr =  2 \cos i \frac{h}{(h^2+r_{\rm{NS}}^2)^{1/2}},
\end{equation}

\begin{figure}
\centering
\includegraphics[width=\columnwidth]{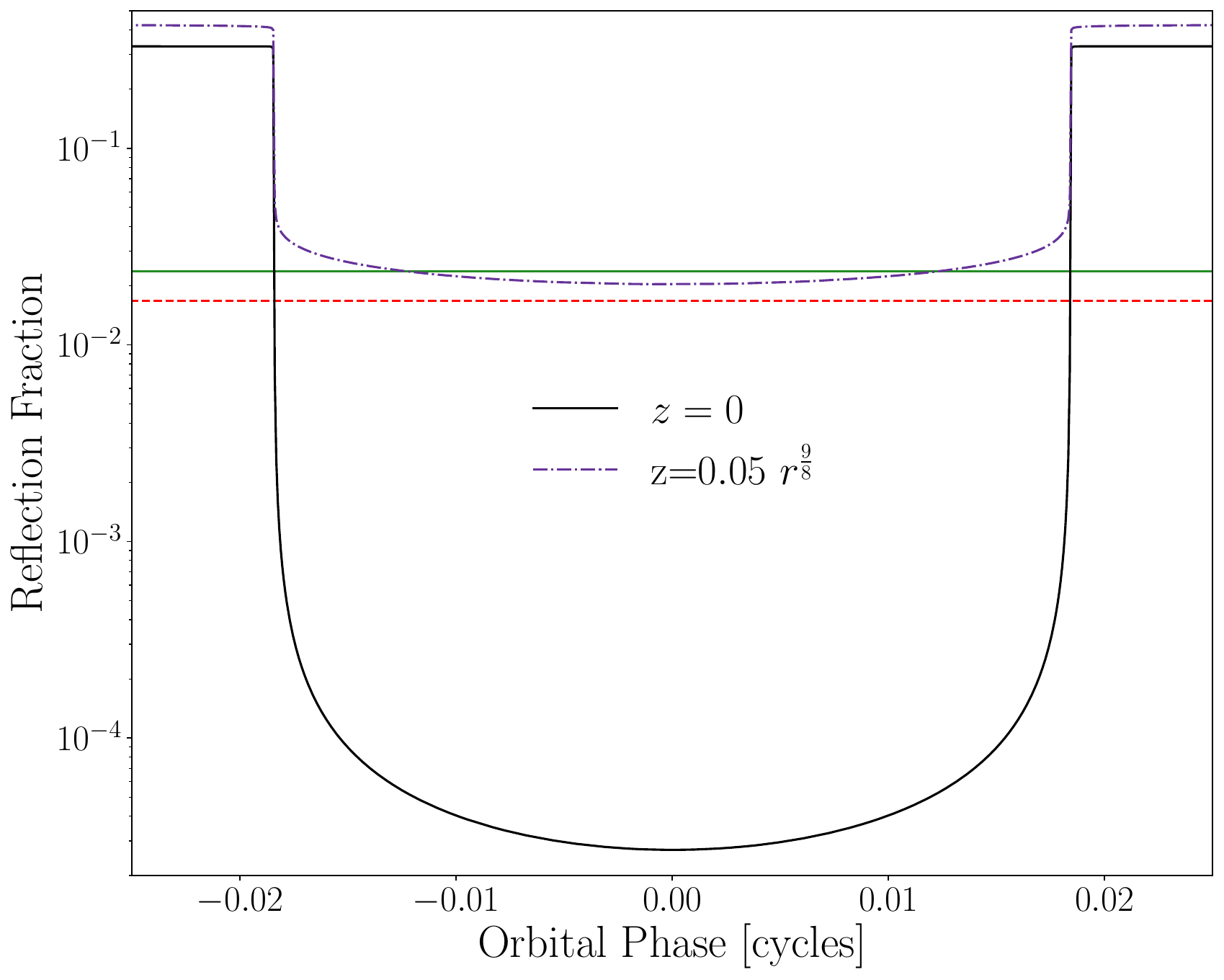}
\vspace*{-0.5cm}
\caption[Ratio of Reflected Flux to Out-Of-Eclipse Direct Flux for a Flared Disc]{Ratio of reflected flux to out-of-eclipse direct flux as a function of orbital phase for a disc being illuminated by a lamppost source at $h=r_{\rm ns}$. For the solid line, the disc is flat ($z=0$) and for the purple, dot-dashed line, it is flared ($z\propto r^{9/8}$). The solid horizontal line indicates the observed $2.4\%$ ratio of in-eclipse burst peak flux to out-of-eclipse burst peak flux and the horizontal dashed line depicts the $1.67\%$ quiescent reflection fraction.
}
\label{fig:ratio_flat}
\end{figure}

For a partially visible accretion disc, we evaluate the above integral numerically by first computing the flux contributed by each surface area element on a grid of $r$ and $\phi$ (disk azimuth) values, $dF(r,\phi) = (dF/dr)~dr~d\phi/(2\pi)$.
For each point on the grid, we calculate the projected distance on the observer's image plane between the centre of the disc patch and the centre of the companion star. If this distance exceeds the radius of the companion star, we include the flux from that patch in the integral. We repeat the calculation for a range of orbital phases and show the results with the solid line in Figure \ref{fig:ratio_flat}. 
 
We see that, at the centre of totality, the reflected emission reaching the observer is $\approx 3 \times 10^{-5}$ times the out-of-eclipse flux and at other orbital phases, the companion star does not cover as much of the accretion disc so we see a greater fraction of the X-ray flux. Overall, this simple model predicts that the bursts peaking during totality should have a peak flux $\sim 3 \times 10^{-5}-10^{-4}$ times the peak flux of the out-of-eclipse bursts. However, the observed value is $\approx 2.4\%$, so reflection from the outer edge of a flat accretion disc is unable to explain the observations of the in-eclipse X-ray bursts found with \textit{RXTE}. 

\begin{figure}
\centering
\includegraphics[width=0.8\columnwidth,trim=0.0cm 7.0cm 0.0cm 1.0cm,clip=true]{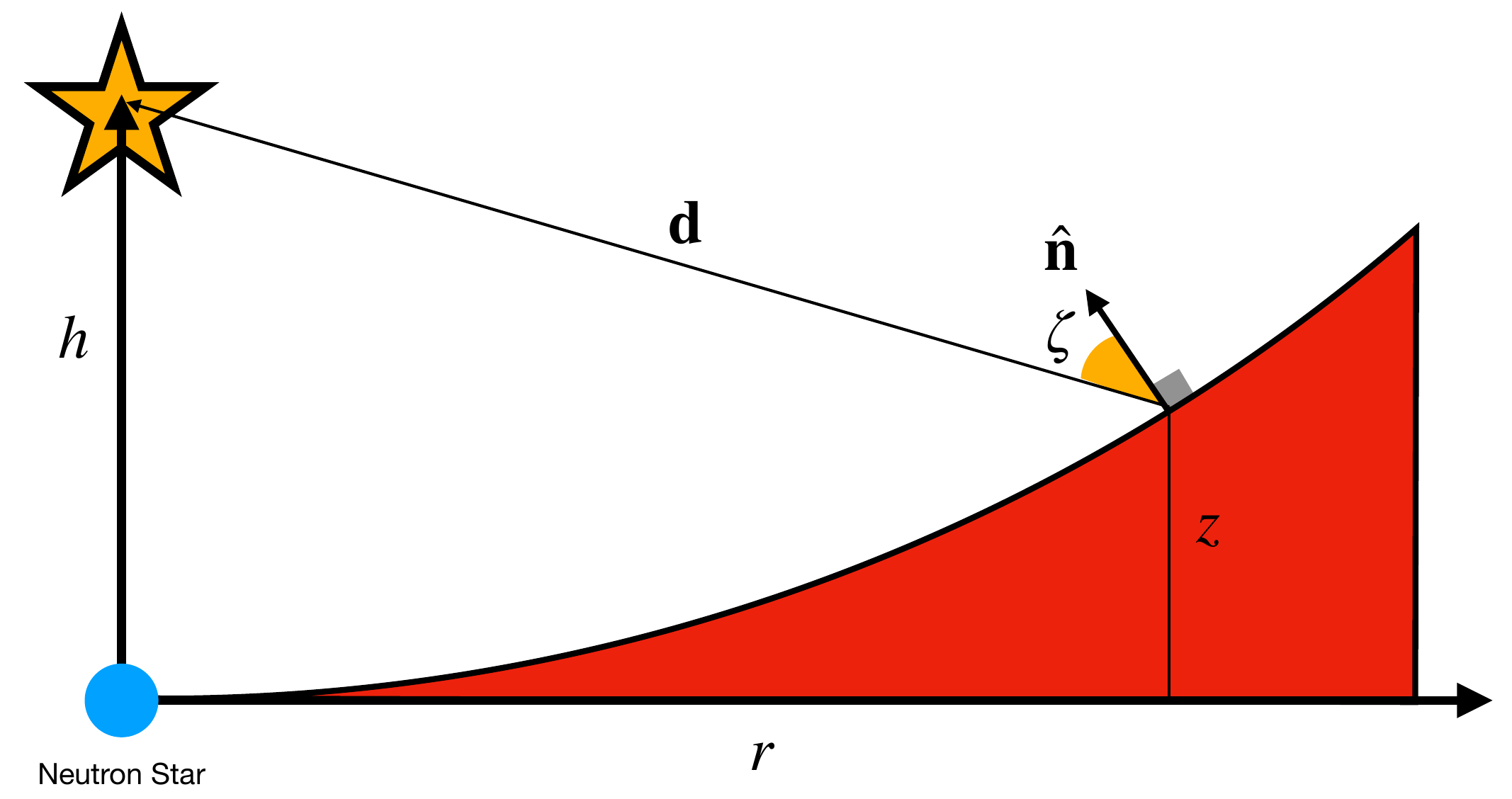}
\vspace{0mm}
\caption[Schematic of a flared disc being illuminated by a lamppost source]{Schematic of a flared disc being illuminated by a lamppost source. A given disc patch is at height $z(r)$ and cylindrical polar radius $r$. The illuminating flux depends on the distance $d$ from the source to the patch and the angle $\zeta$ between the vector $\mathbf{d}$ and the disc normal, $\mathbf{\hat{ n \\}}$.}
\label{fig:flare1}
\end{figure}

Next, we determine the reflection fraction for the case of a flared accretion disc, with a height $z(r)$, as is illustrated in Figure \ref{fig:flare1}. In this case, Equation (\ref{eqn:dfdr_flat}) generalises to

\begin{dmath}
    \frac{dF}{dr~d\phi} = \frac{ r ~{\rm MAX}\left\{\left( r ~dz/dr + h - z \right) , 0 \right\} {\rm MAX}\left\{\left( \cos i - \sin i ~\cos\phi ~dz/dr \right) , 0 \right\} } { \pi \left[ r^2 + (h-z)^2 \right]^{3/2} \left[ (dz/dr)^2 + 1 \right]^{1/2}}.
    \label{eqn:dfdr_flare}
\end{dmath}

Here, the disc azimuthal angle $\phi$ is zero for disc patches that lie in the unique plane defined by the observer's line of sight and the disc rotation axis, and $r$ is the \textit{cylindrical} radial polar coordinate of the disc. We show the result of this calculation with the dot-dashed line in Figure \ref{fig:ratio_flat} and compare it to the case of a flat accretion disc. Assuming a maximally flared \footnote{If the disc were more flared then the outer edges of the disc would completely block our view of the NS. However, \citet{Keek2016} found that for EXO 0748--676 things are more complex. They suggest that the accretion disc may change shape with spectral state, and could even evolve during an X-ray burst.} accretion disc, $z(r) = 0.05 r^{9/8}$, where the power-law index is consistent with zone C of the \citet{Shakura1973} disc model, the out-of-eclipse reflected flux is only just larger than the flat disc case ($0.35$ vs $0.33$), since the total reflected flux is dominated by reflection from the inner disc, which is very thin in both cases. However, the in-eclipse reflected flux is higher than for a flat disc, increasing to $\sim 2\%$, which is closer to the observed value of $\approx 2.4 \%$ and would be sufficient to explain the presence of X-ray bursts peaking during totality for some cases in Figure \ref{fig:refhist}. This increase is because the edge of the outer disc now subtends a larger solid angle to both the observer and the lamppost source. Therefore, we suggest that reflection off of the outer accretion disc likely contributes some of the flux during eclipses, but cannot be entirely responsible for the presence of all X-ray bursts peaking during totality.  

\subsection{Reflection by the Ablated Material}
\begin{figure}
    \centering
    \includegraphics[width=\columnwidth]{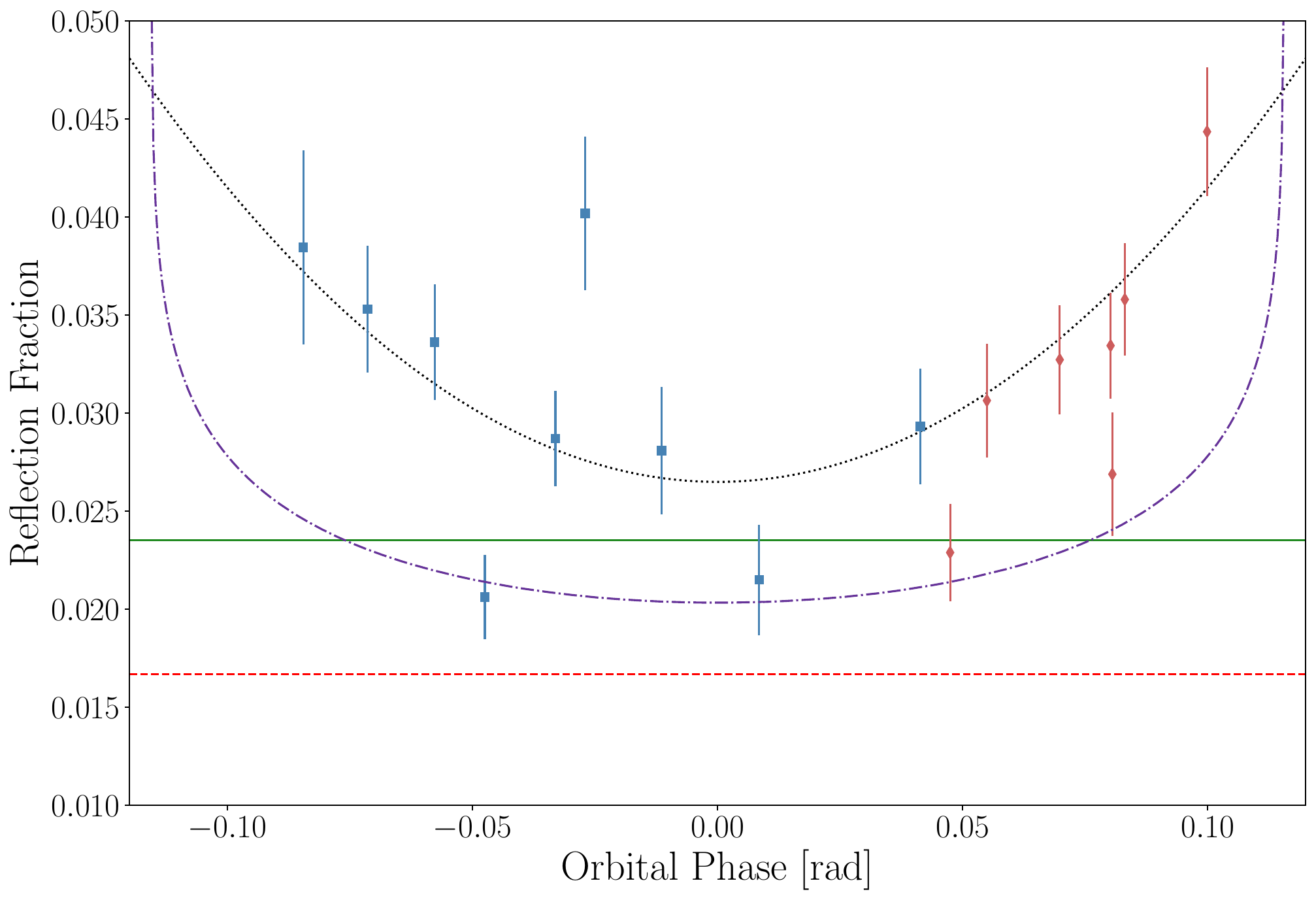}
    \vspace*{-0.5cm}
    \caption[Peak Count Rate of In-Eclipse X-ray Bursts as a Function of Orbital Phase]{The peak count rate of in-eclipse (blue, squared) and egress split bursts (red, diamonds) as a function of orbital phase. Here the orbital phase, $0.0$, marks the centre of totality. Further from the centre of totality, the bursts appear to reach higher peak count rates as demonstrated by the dashed, black trend-line. The solid and dashed horizontal lines show the mean reflection fraction, $0.024$, of the in-eclipse bursts and the quiescent reflection fraction, $0.0167$, respectively. The purple dot-dashed line represents reflection by a maximally flared accretion disc, reproduced from Figure \ref{fig:ratio_flat} for comparison.}
    \label{fig:burst_phase}
\end{figure}

\begin{figure*}
    \centering
    \includegraphics[width=\textwidth]{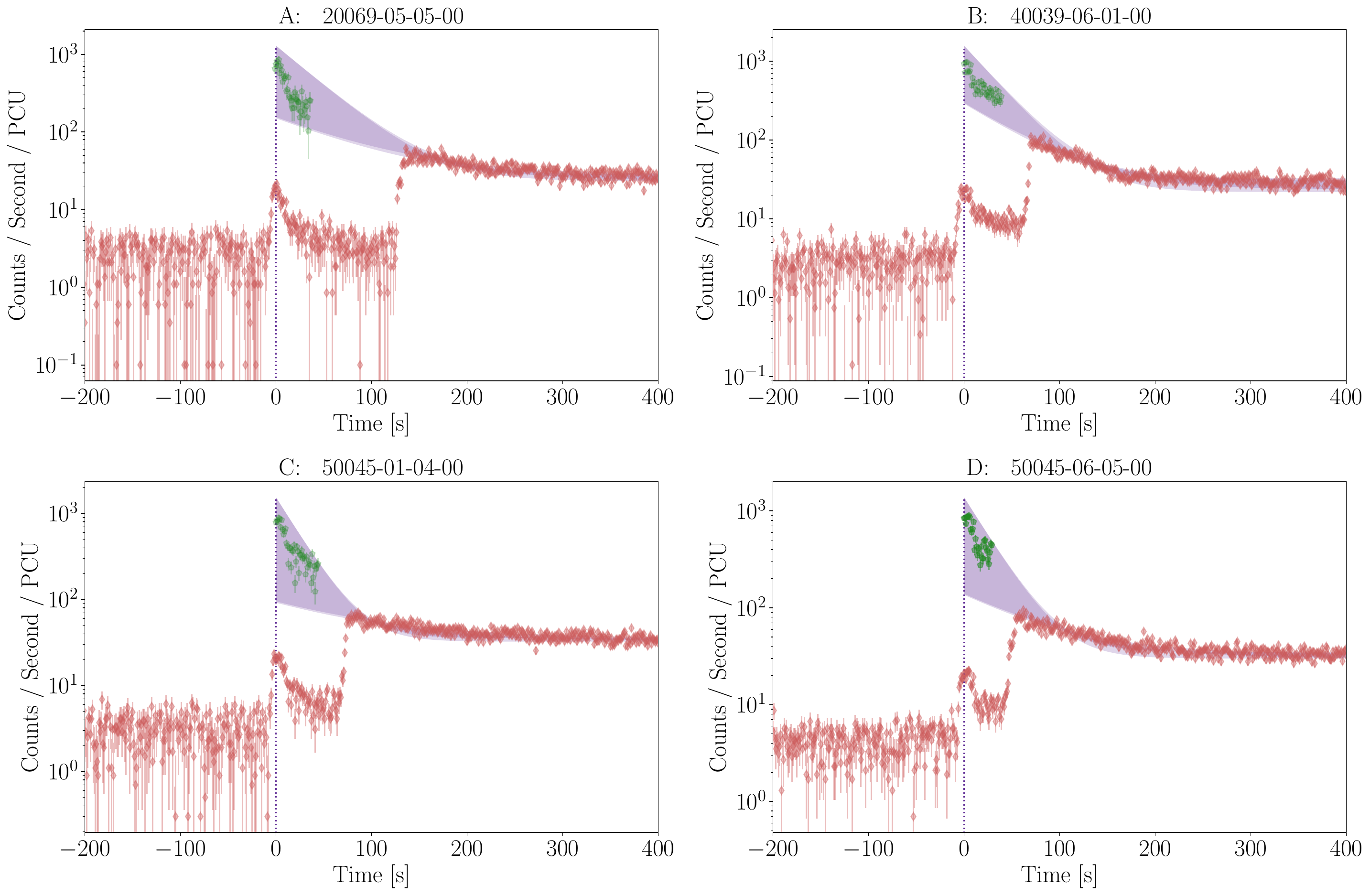}
    \vspace*{-0.5cm}
    \caption{Fast rise -- exponential decay fits to the part of the decay tail that occurs during out-of-eclipse phases for four egress split bursts, labelled in each panel. The fit is extrapolated to the peak of the burst (normalised such that it occurs at zero seconds). Many different FRED profiles returned similar fit statistics, thus we show the range of possibilities through the purple shaded region on each panel. The green data show the in-eclipse peak, multiplied up using the average calculated reflection fraction.}
    \label{fig:FRED}
\end{figure*}

EXO 0748--676 is an archetypal false widow binary, known to undergo irradiation-driven ablation, leading to clumpy, highly ionised material to accumulate close to the companion star. X-ray analysis indicates that the material is highly ionised, $\log(\xi) \gtrsim 3.0$ and extends $\sim 700 - 1500$ km from the stellar surface \citep{Knight2023}. Ablated material can move around the system over time \citep{Polzin2019} and the previously measured variations in density and covering fraction of the liberated stellar material indicates that the material breaks up into clumps as it gets further from the companion (see Section 3 of \citealt{Knight2023} who measured changes in both the density and covering fraction of the ablated material around EXO 0748 on $\sim$ minute timescales). Furthermore, recent optical analysis of the false widow candidate \textit{Swift} J1858.6--0814, suggests that the ablated material could extend much further around the orbit than the X-ray analysis implies \citep{Rhodes2025}. Therefore, ablated material may also play an important role in the reprocessing of X-ray burst emission. Let us consider a scenario whereby the in-eclipse and egress-split bursts arise due to reflection/scattering off this ablated material (or, more accurately, reflection off of the ablated material dominates over reflection off of the outer disc, which we expect to also be present).

The ablated material governs the shape and duration of the ingress and egress by gradually absorbing the X-rays emitted from the NS and accretion disc, typically causing extended eclipse transitions \citep{Knight2022a,Knight2023}. Therefore, if the ablated outflow is the reflector, we hypothesise that the peak count rate of the bursts occurring closer to an eclipse transition will be higher than those coinciding with the centre of totality. To explore this, we calculate the individual reflection fractions, following the same Monte-Carlo procedure as described in Section \ref{sx:Stats} to account for the variations in the out-of-eclipse burst population, for all bursts peaking during totality and plot these as a function of orbital phase in Figure \ref{fig:burst_phase}. Here, we see evidence for our hypothesis; $\phi = 0.0$ rad denotes the centre of totality and the reflection fraction increases on either side as shown by the black dashed trend line. This suggests that the reflection fraction, and therefore peak flux of the burst, is influenced by the the gradual absorption of the X-rays by the ablated outflow. However, it is known that in systems undergoing ablation, the material trails behind the companion star due to the orbital motion of the system \citep{Fruchter1988, Polzin2019, Knight2022a, Knight2023}. As a result, asymmetric orbital phase dependence is predicted if the ablated outflow is the reflection site, although calculating the expected degree of asymmetry is beyond the scope of this paper. Thus, the symmetric phase dependence found in Figure \ref{fig:burst_phase} could simply be informing us that more of the scatterer is visible either side of the eclipse. Interestingly, Figure \ref{fig:burst_phase} also shows that many of the bursts peaking in totality have individual reflection fractions higher than the $2.4 \%$ calculated in Section \ref{sx:Stats}. This is unlikely to be caused by the reflecting site itself, regardless of what site that may be, and instead favours a scenario whereby the bursts themselves are increasing the amount of scattering material (e.g. \citealt{Albayati2023}). We also suggest that more than one reflection site may be responsible for the observed population of bursts peaking during totality. Figure \ref{fig:burst_phase} clearly shows a subset of four bursts that are consistent with reflection by a flared accretion disc, while the rest of the population require a different reflection site. As such, we cannot rule out multiple/changing reflection sites.

We further explore the phase dependence of the reflection fraction by performing a series of exponential decay fits to the out-of-eclipse tail of the egress-split bursts to ascertain the brightness of their peaks, had they occurred during an out-of-eclipse phase. Figure \ref{fig:FRED} shows these fits for four of the brightest egress-split bursts, where the dark purple shaded region depicts the $3 \sigma$ region for all possible fits with $\chi^{2}_{\nu} \leq 2.0$. We find that a range of exponential decay equations can fit each out-of-eclipse tail, making it challenging to determine how bright each burst would have been but we do find that our calculated mean reflection fraction of $2.4 \%$ is consistent in all four cases. We show this consistency by multiplying up the peak of the burst accordingly and plot these with green pentagons in each panel of Figure \ref{fig:FRED}. The range of possible out-of-eclipse peaks predicted by our fits enables an independent measurement of the reflection fraction for each case in Figure \ref{fig:FRED}. For ObsID 20069-05-05-00, the range of peaks predicted by our fits is $152.95 - 1293.26$ cts/s/PCU corresponding to reflection fractions in the range $0.0165 - 0.140$. Similarly for ObsIDs 40039-06-01-00, 50045-01-04-00 and 50045-06-05-00, respectively, the range of peaks is $292.42 - 1568.05$, $93.59 - 1546.32$ and $137.75 - 1382.75$ which correspond to reflection fractions in the ranges $0.0159 - 0.0852$, $0.0151 - 0.249$ and $0.0157 - 0.157$. Thus, it is possible that the reflection fraction is significantly higher than our statistics determine. However, the lower limit in the reflection fraction in all cases is less than the reflection fraction of the quiescent emission, suggesting that our understanding of these events is incomplete. Thus further observations of these rare events are essential to understanding their origin. 

Finally, we note a few pieces of earlier evidence that assist the argument for reflection by the ablated outflow. Firstly, our spectral analysis implies that the accretion disc is not significantly contributing to the spectra of the bursts. While this is likely due to the unocculted outer disc being too cool to detect in the \textit{RXTE} spectra, it is also explainable if the ablated material extends further from the companion star or moves significantly (see e.g. \citealt{Knight2023}). Furthermore, the spectral models that include reflection components prefer high ionisations, consistent with the previously measured ionisation of the ablated material \citep{Knight2022a}. It is also important to note that the presence of a significant amount of ablated material may explain the lack of in-eclipse X-ray bursts observed by \textit{XMM-Newton} since the softer X-rays that \textit{XMM-Newton} are sensitive to will be heavily absorbed by the ablated material. However, we note that other factors including the observing mode, detector sensitivity and a higher residual in-eclipse emission than \textit{RXTE} may also explain the lack of X-ray bursts observed by \textit{XMM-Newton}. Lastly, the ablated material provides a way to explain the lack of observed X-ray bursts between $53500 - 54000$ MJD, as this coincides with the eclipse asymmetry reversing (due to the movement of the ablated material around the system) and a probable spectral state change \citep{Knight2023}. We speculate that this may be because the ablated material obscures our view of the bursts, or that the material was no longer suitably positioned to reflect the burst emission. Although, this may simply be due to a change in the accretion rate, and therefore burst recurrence rate, caused by the change in spectral state. 

\subsection{Reflection by an accretion disc wind}
\subsubsection{Simulations}
The final possibility we consider is that the burst photons are scattered around the eclipsing companion by an accretion disc wind -- an interesting prospect given the detection of Fe K$\alpha$ absorption lines in this system by \cite{Ponti2014}. To investigate this possibility, we conducted Monte Carlo radiative transfer (MCRT) simulations to model Compton scattering in an equatorial biconical disc wind. We use the MCRT code {\textsc{Sirocco}} \citep{long_modeling_2002,matthews2024} for this purpose, which assumes an azimuthally symmetric wind but tracks photon trajectories in 3D. The code also allows the inclusion of a Roche-lobe filling secondary star which is treated as a purely absorbing surface. The wind itself is parametrized according to the prescription of \cite{shlosman_winds_1993}, which has been used regularly to model accretion disc winds. In the context of X-ray binaries, we take as our starting point the wind geometry used by \cite{koljonen2023} to simulate the optical spectrum of MAXI J1820$+$070 in the soft state. Our focus in this case is a ``pure scattering'' scenario; thus, rather than doing a full photoionization calculation, we simply adopt Saha abundances and set the wind electron temperature to $T_e=10^7{~\rm K}$ so that the wind is mostly ionized. We launch photon packets isotropically from the neutron star surface (neglecting GR effects), and the calculation does not depend on the assumed spectrum in this pure scattering limit. We then compute synthetic spectra at close to the system inclination ($i=77^\circ$), both in-and-out of eclipse. By comparing the observed in-and-out of eclipse luminosities, we can calculate the reflection fraction. 

We conducted a small grid of simulations with a fixed wind geometry (wind opening angles of $\theta_{\rm min} = 60^\circ$ to $\theta_{\rm max} = 89.9^\circ$) and other wind parameters chosen as in \cite{koljonen2023}, except for the volume filling factor which was set to one (the results are insensitive to this latter choice). We varied the total mass-loss rate of the wind, $\dot{M}_{\rm wind}$, and the inner wind launch radius $r_{\rm min}$. We also fixed the outer wind launch radius, $r_{\rm max} = 10/7~r_{\rm min}$. All of these parameters are somewhat unconstrained; however, our aim here is not to conduct a full parameter search or fit, but instead demonstrate the plausibility (or lack thereof) of a wind scattering scenario. 

\begin{figure}
\centering
\includegraphics[width=\columnwidth]{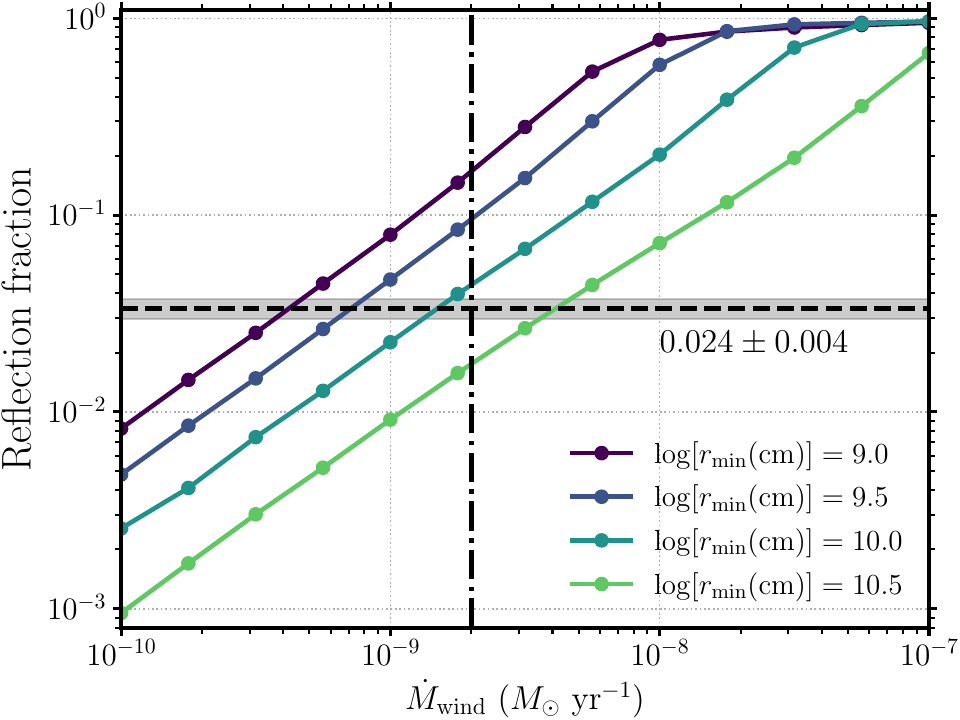}
\vspace*{-0.5cm}
\caption{Reflection fraction $R$ from a biconical, equatorial accretion disc wind, obtained from Monte Carlo radiative transfer simulations (using \textsc{Sirocco}) in an azimuthally symmetric geometry with an eclipsing companion. Each coloured line connects points showing results from individual simulations with a given mass-loss rate, $\dot{M}_{\rm wind}$, and inner launch radius, $r_{\rm min}$. The horizontal dashed line marks the mean  reflection fraction of $F_{\rm{reflect}} = 0.024$ and  corresponding standard deviation of $0.004$ required to explain the in-eclipse bursts, and the vertical dot-dashed line marks $\sim 2\dot{M}_{\rm acc}$, a reasonable value for $\dot{M}_{\rm wind}$.}
\label{fig:wind}
\end{figure}

Fig.~\ref{fig:wind} shows the in-eclipse reflection fraction from our MCRT simulations, as a function of $\dot{M}_{\rm wind}$, for four different values of $r_{\rm min}$. The horizontal dashed line marks the mean value of $F_{\rm reflect}$, the reflection fraction required to explain the in eclipse bursts. The reflection fraction increases with $\dot{M}_{\rm wind}$, because the optical depth of the wind increases, and also increases as the launching radius is moved closer to the neutron star, because the wind intercepts more radiation. For a far out wind, $\log [r_{\rm min} ({\rm cm}) = 10.5$, we require $\dot{M}_{\rm wind}\approx 3\times10^{-9}~M_\odot~{\rm yr}^{-1}$ to reproduce the required $F_{\rm reflect}$, whereas a close-in wind, $\log [r_{\rm min} ({\rm cm})] = 9$, can do so for $\dot{M}_{\rm wind}\approx 3\times10^{-10}~M_\odot~{\rm yr}^{-1}$. This compares to an approximate accretion rate of $\dot{M}_{\rm acc}\sim 10^{-9}~M_\odot~{\rm yr}^{-1}$ based on a previously measured out-of-burst luminosity of $6\times10^{36}~{\rm erg~s}^{-1}$ \citep{Sidoli2005,Boirin2007} combined with an assumed radiative efficiency of $\eta = 0.1$. We might expect $\dot{M}_{\rm wind} \sim 2 \dot{M}_{\rm acc}$ in typical XRBs \citep{Ponti2012,higginbottom19}, which we mark with a vertical dot-dashed line. Our estimates therefore suggest that the mean burst reflection fraction of $2.4\%$ can easily be produced by an accretion disc wind for plausible mass-loss rates and launching radii. For close-in winds or high mass-loss rates, higher reflection fractions of $\gtrsim 10 \%$ are feasible.

\subsubsection{Discussion}
Our calculations show that a reasonable biconical wind structure can produce the observed scattering fraction, but this model is not yet predicated on a specific wind-driving mechanism. It is thus important to consider which mechanisms -- if any -- could drive the putative accretion disc wind. We first consider Compton-heated or thermal winds, which can be explored using the framework provided by \cite{Begelman_compton_1983} and updated by \cite{Done_thermal_2018}. A natural scenario to test is one where a thermal wind is ever-present and responsible for a scattering contribution that explains the observed in-eclipse continuum level, with a geometric effect or enhanced wind power that can reproduce our measured reflection fraction, and our finding of $F_{\rm reflect}>R$.

The key quantities for determining whether a thermal wind can be launched are the Compton radius and the critical luminosity. The former is given by
\begin{equation}
r_{\rm IC} \approx 2\times 10^{11}~
\left(\frac{M}{2M_\odot}\right) 
\left(\frac{10^7~{\rm K}}{T_{\rm IC}}\right)~{\rm cm}
\end{equation}
where $T_{\rm IC}$ is the Compton temperature and $T_{\rm IC}\approx 10^7~{\rm K}$ for the non-bursting/persistent emission spectral energy distribution (SED). The critical luminosity is given by 

\begin{equation}
L_{\rm crit} \approx 0.094
\left(\frac{T_{\rm IC}}{10^7~{\rm K}}\right)^{-1/2}~L_{\rm Edd}.
\end{equation}
To drive a thermal wind, one requires both $L>L_{\rm crit}$ and $r_{\rm disc} > 0.2~r_{\rm IC}$, where $r_{\rm disc}$ is the maximum extent of the accretion disc. The mass-loss rate will also rise with an increasing disc size \citep{Done_thermal_2018}, since it has a larger reservoir of mass that can be heated to the escape speed. The disc size can be estimated, following, e.g., \cite{Eggleton1983,DiazTrigo_accretion_2016,Mushtukov2019}, as $r_{\rm disc} \approx 0.9 r_L$, where $r_L$ is the size of the Roche lobe. This gives an estimate as $r_{\rm disc} \approx 4.5 \times 10^{10}~{\rm cm} \sim 0.2 r_{\rm IC}$. Again taking $6\times10^{36}~{\rm erg~s}^{-1}$ as a typical out-of-burst luminosity we find $L \approx 0.25 L_{\rm crit}$.

There are thus two problems with driving a thermal wind in EXO 0748 outside of the bursts: the disc is marginally too small (or right on the cusp of the critical radius), and the luminosity is too low; the former issue was noted by \cite{DiazTrigo_accretion_2016}, who suggest the Fe K absorption found by \cite{Ponti2014} is associated with an `atmosphere' rather than an outflow. During the bursts, the luminosity increases dramatically, and the softening of the SED also leads to a decrease in $T_{\rm IC}$. These effects result in moving upwards and left in the \cite{Begelman_compton_1983} diagram, likely leading to a Compton-heated atmosphere or corona, but not a successful thermal wind. This means, in fact, that EXO 0748 is a particularly good laboratory for disc wind driving mechanisms, for two main reasons: i) if a reliable outflow velocity could be associated with the soft-state Fe K$\alpha$ absorption it could act as compelling evidence for a magnetohydrodynamic (MHD) wind origin; and ii) it in principle allows the response of any disc wind to irradiation and heating by the burst to be studied.

With a Compton-heated thermal wind unlikely, we are left with a few possible options. One is that there is indeed an MHD wind in the system, which scatters the burst radiation into our line of sight. MHD winds have been invoked in GRO 1655$-$40 based on inferred launching radii (\citealt{miller2008}; although see also \citealt{Tomaru2023_GRO1655}). A second possibility is that the wind is actually radiatively driven by the burst itself, given that the luminosity increases to near Eddington values. This scenario is compelling, but would require an alternative (i.e., non-wind) explanation for the reflected continuum during the eclipse. A final option is that the Compton-heated atmosphere has sufficient scale height, or works in tandem with flaring in the disc, to scatter around the companion star. The distinction between these three possibilities is somewhat blurred and testing them is challenging without further disc wind modelling and high-resolution X-ray spectroscopy. Nevertheless, our overall findings are that a reasonable biconical disc wind geometry and mass-loss rate can naturally produce the required reflection fraction for the in-eclipse bursts. 

A complete disc wind model for the in-eclipse burst phenomenology should also explain our finding of $F_{\rm reflect}>R$. In the MHD wind setting, for example, this would require that the wind was a more effective scatterer during the burst. A denser wind could be feasible if there is additional heating in the launching region; for example, \cite{casse2000, Ferreira2004, Chakravorty_magneto_2016} find warm MHD solutions are more mass-loaded than cold solutions. However, any dynamic changes in the wind would have to take place extremely quickly, either in advance of, or over the duration of, the burst, for this explanation to work, which is a challenge when considering, e.g., the flow time of a disc wind. It is perhaps more likely that $F_{\rm reflect}$ is higher than $R$ because of the anisotropy of the radiation field during the burst; that is, the burst radiation has a radiation pattern such that the scatterer intercepts more burst radiation relative to the out-of-burst continuum. Alternatively, the in-eclipse continuum could simply be produced (or scattered) by a different component compared to the in-eclipse burst. 

\vspace*{-1cm}
\section{Conclusions}
\label{sx:Con}
In this paper, we present 22 Type I X-ray bursts observed by \textit{RXTE} that all coincide, fully or partially, with an X-ray eclipse, exhibited by the false widow binary EXO 0748--676. We identify nine instances where the burst occurs entirely within totality, seven bursts split across an egress, and six cases interrupted by an ingress. All in-eclipse bursts and split bursts occurred while the source was in the hard spectral state and we deem that all presented events are physical based on their PCU behaviour. 

We determine that we are not observing direct burst emission during the eclipses and thus assess several scenarios whereby the burst emission is reflected/scattered into our line-of-sight, thus allowing us to observe them. We conclude that reflection off of a flat accretion disc is not plausible as the reflection fraction achieved is insufficient to be distinguished from the residual flux level during the X-ray eclipses. However, the reflection fraction achieved by a maximally flared accretion disc, $\sim 2~\%$ is consistent with the measured reflection fraction of $0.024 \pm 0.004$ ($1~\sigma$ uncertainty). Therefore, a flared disc could explain these observations in some cases. In particular, we note that four of the bursts that peak during totality are fully consistent with a maximally flared disc as the reflection site. However, the remaining $12$ bursts that peak during totality require an alternative explanation. Therefore, it is possible that the bursts observed to peak during eclipses are reflected by different components of the system. We find the reflector/scatterer must subtend a large solid angle in order to meet the observed reflection fractions. Therefore, scenarios whereby the bursts are reflected off an accretion disc wind or the ablated material known to surround EXO 0748--676 are plausible. Furthermore, all in-eclipse and split X-ray bursts occurred while the source was in the hard spectral state, so reflection by an accretion disc corona also cannot be ruled out. 

In summary, we cannot confidently determine the reflection site for all X-ray bursts peaking during totality. Observations of in-eclipse bursts from EXO 0748--676 that leverage both the energy and time domains, would allow us to measure the effective area and energy response of the reprocessing material and thus distinguish between the proposed scenarios. Observations with \textit{NICER}, while EXO 0748--676 is in outburst, are ideal for this purpose and would provide the high time- and spectral- resolution needed to separate the persistent and reflected burst emission during eclipses.

\section*{Acknowledgements}
The authors acknowledge valuable discussions with A. C. Albayati, who provided useful insights regarding future observations of in-eclipse X-ray bursts. AHK acknowledges support from the Oxford Hintze Centre for Astrophysical Surveys, which is funded through generous support from the Hintze Family Charitable Foundation. TPR and AHK acknowledge support from the Science and Technology Facilities Council (STFC) as part of the consolidated grant award ST/X001075/1. AI and JHM acknowledge support from the Royal Society. AI and JHM acknowledge support from the Royal Society. MM acknowledges support from the Science and Technology Facilities Council (STFC) as part of the consolidated grant award ST/V001000/1. GCM was partially supported by PIP 0113 (CONICET). The authors thank the anonymous referee for the valuable feedback that improved this paper.

\section*{Data Availability}
The data used in this study are publicly available from the HEASARC website or the \textit{XMM-Newton} Science Archive. This study has made use of software and tools provided by HEASARC and ESA. The \textit{RXTE} burst index will be available online as with this paper as supplementary material.



\bibliographystyle{mnras}
\bibliography{All_Refs} 

\begin{appendices}
\section{Data Tables}

\onecolumn
\begin{table*}
\caption{\label{tb:RXTE_Table} \textit{RXTE} X-ray burst index for EXO 0748 -- 676. We note if the burst was previously identified in \citealt{Galloway2008} (GW08) or \citealt{Wolff2009} (W09) and note the Multi-INstrument Burst ARchive (MINBAR; \citealt{Galloway2020}) record number where applicable. The full version of this table is available online as supplementary material.}
\begin{tabular}{c c c c c c c c c}
\hline
ObsID & Orbital & No. & Peak Burst & Peak Burst & Peak Count  & Classification & MINBAR & Additional \\ \vspace*{0.075cm}
& Cycle & PCU & MJD & Count Rate & Rate Error &  & Record No. & References \vspace*{0.075cm} \\ \hline 
10108-01-06-00 & 26353 & 5 & 50310.10126 & 57.5 & 5.99 & In-Eclipse & & \vspace*{0.075cm} \\
10108-01-07-01 & 26358 & 5 & 50310.90805 & 3273.9 & 56.9 & Out-of-Eclipse & 2237 & GW08  \vspace*{0.075cm}  \\ 
10108-01-10-00 & 26356 & 4 & 50310.73536 & 886.6 & 29.4 & Out-of-Eclipse & 2235 & GW08  \vspace*{0.075cm} \\
10108-01-10-00 & 26357 & 5 & 50310.74719 & 2824.5 & 52.9 & Out-of-Eclipse & 2236 & GW08  \vspace*{0.075cm} \\
10108-01-12-00 & 26655 & 5 & 50358.25337 & 5309.7 & 72.7 & Out-of-Eclipse & 2240 & GW08 \vspace*{0.075cm} \\ 
10108-01-13-00 & 26656 & 5 & 50358.39302 & 3874.7 & 62.0 & Out-of-Eclipse & 2241 & GW08 \vspace*{0.075cm} \\
20069-04-03-00 & 27983 & 5 & 50569.83106 & 6606.0 & 79.2 & Out-of-Eclipse & 2288 & GW08  \vspace*{0.075cm} \\
20069-05-05-00 & 28335 & 4 & 50625.91172 & 85.4 & 8.14 & Egress-Split & & W09 \vspace*{0.075cm} \\
20082-01-01-00 & 28636 & 5 & 50673.96185 & 4831.0 & 68.6 & Out-of-Eclipse & 2334 & GW08  \vspace*{0.075cm} \\
20082-01-02-00 & 28660 & 5 & 50677.63089 & 3195.0 & 56.1 & Out-of-Eclipse & 2335 & GW08   \vspace*{0.075cm} \\
20082-01-02-000 & 28662 & 5 & 50678.08744 & 5815.1 & 102 & Out-of-Eclipse & 2336 & GW08  \vspace*{0.075cm} \\
30067-04-03-00 & 30999 & 5 & 51050.38430 & 1350.3 & 36.3 & Ingress-Split & & W09 \vspace*{0.075cm} \\
30067-08-01-00 & 29967 & 5 & 50885.97391 & 5017.5 & 70.6 & Out-of-Eclipse & 2378 & GW08   \vspace*{0.075cm} \\
30067-09-01-00 & 30637 & 5 & 50992.63513 & 5638.5 & 74.9 & Out-of-Eclipse Doublet & 2395 & GW08 \vspace*{0.075cm} \\
30067-09-01-00 & 30637 & 5 & 50992.64466 & 4720.5 & 68.5 & Out-of-Eclipse Doublet & 2396 & GW08 \vspace*{0.075cm} \\
30067-12-01-00 & 31669 & 4 & 51157.26329 & 3368.2 & 57.8 & Out-of-Eclipse & 2464 & GW08 \vspace*{0.075cm} \\
40039-03-04-00 & 32628 & 3 & 51309.93793 & 3178.2 & 56.1 & Out-of-Eclipse & 2531 & GW08  \vspace*{0.075cm} \\
40039-04-04-00 & 32968 & 4 & 51364.12049 & 97.2 & 8.53 & In-Eclipse & 2550 & GW08, W09 \vspace*{0.075cm} \\
40039-04-05-00 & 32969 & 3 & 51364.27694 & 1721.5 & 41.3 & Ingress-Split & 2551 & GW08, W09 \vspace*{0.075cm} \\
40039-05-02-00 & 33265 & 4 & 51411.60057 & 1496.2 & 38.3 & Out-of-Eclipse & 2566 & GW08 \vspace*{0.075cm} \\
40039-06-01-00 & 33621 & 5 & 51468.17349 & 124.7 & 10.0 & Egress-Split & 2592 & GW08, W09 \vspace*{0.075cm} \\
50045-01-04-00 & 34651 & 5 & 51632.28938 & 116.5 & 9.42 & Egress-Split & 2614 & GW08,  W09 \vspace*{0.075cm} \\
50045-03-02-00 & 35305 & 5 & 51736.49148 & 3003.5 & 54.5 & Ingress-Split & 2634 & GW08, W09 \vspace*{0.075cm} \\
50045-03-05-00G & 35309 & 4 & 51737.12942 & 2906.3 & 52.8 & Ingress-Split & 2635 & GW08, W09 \vspace*{0.075cm}\\
50045-04-03-00 & 35612 & 5 & 51785.41090 & 100.0 & 8.51 & In-Eclipse & 2672 & GW08 \vspace*{0.075cm} \\
50045-06-01-00 & 36305 & 4 & 51895.84346 & 2997.4 & 54.6 & Out-of-Eclipse & 2696 & GW08  \vspace*{0.075cm} \\
50045-06-02-00 & 36306 & 5 & 51895.99542 & 154.5 & 11.4 & Egress-Split & 2697 & GW08, W09 \vspace*{0.075cm} \\
50045-06-05-00 & 36310 & 5 & 51896.63246 & 108.3 & 9.21 & Egress-Split & 2698 & GW08, W09 \vspace*{0.075cm} \\
\bottomrule 
\end{tabular}
\end{table*}

\newpage
\begin{table}
\centering
\begin{tabular}{c c c c c c c}
\hline
ObsID & Event No. & Event MJD & \quad & ObsID & Event No. & Event MJD\\
\hline
30067-06-04-00 & 1 & 51157.14093 &\quad \quad \quad & 92019-01-20-01 & 1 & 54198.35432 \\
70048-07-01-00 & 2 & 52517.90521 & & 92019-01-20-01 & 2 & 54198.40713 \\
90059-12-04-00 & 1 & 53932.40003 & & 92019-01-22-02 & 1 & 54291.92684 \\
91043-06-04-00 & 1 & 53742.24518 & & 92019-01-22-02 & 2 & 54291.94015 \\
91043-10-02-00 & 1 & 54083.22469 & & 92019-01-22-05 & 2 & 54293.41512 \\
92019-01-01-00 & 1 & 53981.54689 & & 92019-01-25-01 & 1 & 54346.00511 \\
92019-01-03-00 & 2 & 53991.04561 & & 92019-01-26-00 & 1 & 54388.68818 \\
92019-01-09-06 & 1 & 54246.50564 & & 92019-01-26-01 & 1 & 54389.61311 \\
92019-01-09-13 & 1 & 54093.18931 & & 92019-01-28-02 & 1 & 54447.50017 \\
92019-01-10-01 & 1 & 54087.37147 & & 92019-01-33-00 & 1 & 54690.55600 \\
92019-01-12-02 & 1 & 54103.45592 & & 92040-01-03-00 & 1 & 54185.36465 \\
92019-01-14-000 & 1 & 54134.32658 & & 92040-06-02-00 & 1 & 54744.32763 \\
92019-01-15-00 & 1 & 54138.16616 & & 93074-05-01-00 & 1 & 54494.80048 \\
92019-01-15-02 & 1 & 54137.93737 & & 93082-05-01-02 & 1 & 54495.06474 \\
92019-01-16-00 & 1 & 54142.96326 & & 93082-05-01-03 & 1 & 54496.07977 \\
92019-01-18-000 & 1 & 54185.72239 & & 93082-05-01-03 & 2 & 54496.11099\\
92019-01-18-000 & 2 & 54185.86839 & & &  \\
\hline
\end{tabular}
\caption{\label{tb:breakdowns}A list of PCU breakdowns present in the archival \textit{RXTE} data of EXO 0748--676. These events can, visually, resemble bursts but are instrumental. Here, we list the ObsID, the burst/burst-like event number (representing the order of the events in the observations) and the MJD of the event.}
\end{table}

\begin{table*}
\begin{tabular}{c c c c c c}
\toprule
Burst Type & Epoch & T$_{\rm{in}}$ [keV]  & kT [keV] & $\chi/\nu$ & $p$\\
\midrule
In-Eclipse & 3 & 1.3 $\pm$ 0.7 & 0.9 $\pm$ 0.4 & 0.86 & 0.728\\
In-Eclipse & 4 & $> 5.4 $ & 1.4 $\pm$ 0.3 & 1.13 & 0.259 \\
\vspace*{+0.15cm}
In-Eclipse & 5 & 3.3 $\pm$ 0.8 & 1.9 $\pm$ 0.17 & 1.11 & 0.166 \\
Egress-Split (p) & 3 & $> 0.3$ & 1.9 $\pm$ 0.17 & 0.86 & 0.728\\
Egress-Split (p) & 4 & 214 $\pm$ 65 & 1.9 $\pm$ 0.15 & 1.14 & 0.063 \\
\vspace*{+0.15cm}
Egress-Split (p) & 5 & $< 343$ & 1.83 $\pm$ 0.07 & 1.04 & 0.293 \\
Egress-Split (t) & 3 & 2.3 $\pm$ 0.4 & 0.3 $\pm$ 0.2 & 1.00 & 0.465\\
Egress-Split (t) & 4 & 4.7 $\pm 1.3$ & 1.22 $\pm$ 0.07 & 0.99 & 0.491 \\
Egress-Split (t) & 5 & $3.0 \pm 0.3$ & 0.88 $\pm$ 0.15 & 1.22 & 0.111 \\
\end{tabular}

\begin{tabular}{c c c c c c c c}
\toprule
Burst Type & Epoch & N$_{\rm{h}}$ [10$^{22}$ cm$^{-2}$] & f$_{\rm{cov}}$ & $\Gamma$ & $\log(\xi)$ & $\chi/\nu$ & $p$\\
\midrule
In-Eclipse & 3 & $> 149$ & $0.981 \pm 0.007$ & 2.5 $\pm$ 1.1 & 2.0 $\pm$ 1.2 & 0.95 & 0.635\\
In-Eclipse & 4 & 65 $\pm$ 42 & 0.95 $\pm$ 0.04 & 3.3 $\pm$ 1.6 & 2.1 $\pm$ 0.7 & 1.04 & 0.403 \\
\vspace*{+0.15cm}
In-Eclipse & 5 & $> 17.1$ & 0.982 $\pm 0.007 $ & 3.7 $\pm$ 0.8 & 2.9 $\pm 0.3$ & 0.99 & 0.156 \\
Egress-Split (p) & 3 & $> 21.1$ & 0.994 $\pm 0.002$ & 4.6 $\pm$ 0.9 & 2.2 $\pm$ 2.1 & 1.28 & 0.104\\
Egress-Split (p) & 4 & $>14$ & 0.94 $\pm$ 0.06 & 2.1 $\pm$ 0.7 & 2.9 $\pm$ 1.7 & 1.09 & 0.187 \\
\vspace*{+0.15cm}
Egress-Split (p) & 5 & $> 439$ & 0.65 $\pm$ 0.08 & 1.01 $\pm 0.08$ & 3.55 $\pm$ 0.13 & 1.05 & 0.135 \\
Egress-Split (t) & 3 & 73 $\pm$ 56 & 0.89 $\pm$ 0.09 & 2.97 $\pm$ 1.09 & 2.1 $\pm$ 0.7 & 1.04 & 0.364 \\
Egress-Split (t) & 4 & 9.8 $\pm$ 4.5 & $> 0.81$ & 1.85 $\pm$ 0.15 & 1.9 $\pm 1.1$ & 1.47 & 2.2$\times 10^{-6}$ \\
Egress-Split (t) & 5 & $2.1 \pm 0.9$ & 0.82 $\pm 0.18$ & 1.93 $\pm$ 0.14 & $> 1.4$ & 1.45 & 2.9$\times 10^{-6}$ \\
\end{tabular}

\begin{tabular}{c c c c c c c c c}
\toprule
Burst Type & Epoch & N$_{\rm{h}}$ [10$^{22}$ cm$^{-2}$] & f$_{\rm{cov}}$ & kTbb [keV] & $\log(\xi)$ (disc) & $\log(\xi)$ (absorber) & $\chi/\nu$ & $p$\\
\midrule
In-Eclipse & 3 & $> 481 $ & $>0.999$ & 1.03 $\pm$ 0.09 & $< 1.8$ & 2.2 $\pm$ 1.5 & 0.99 & 0.483\\
In-Eclipse & 4 & $> 58$ & $>0.999$ & 1.2 $\pm$ 0.2 & 2.0 $\pm$ 0.3 & 3.5 $\pm$ 3.2 & 0.94 & 0.586 \\
\vspace*{+0.15cm}
In-Eclipse & 5 & 180 $\pm$ 34 & 0.949 $\pm 0.003 $ & 1.5 $\pm$ 0.4 & 2.0 $\pm$ 1.3 & 3.0 $\pm$ 0.8 & 1.03 & 0.372 \\
Egress-Split (p) & 3 & $> 18$ & $> 0.874$ & 0.9 $\pm$ 0.3 & 2.5 $\pm 0.8$ & $>2.2$ & 0.95 & 0.578\\
Egress-Split (p) & 4 & 143 $\pm$ 38 & 0.985 $\pm$ 0.008 & 1.02 $\pm$ 0.07 & $< 1.6$ & $> 2.9$ & 1.01 & 0.538 \\
\vspace*{+0.15cm}
Egress-Split (p) & 5 & $> 242$ & 0.83 $\pm$ 0.06 & 2.00 $\pm$ 0.11 & $< 3.6 $ & 2.7 $\pm$ 1.6 & 1.07 & 0.350 \\
Egress-Split (t) & 3 & $> 78$ & 0.997 $\pm$ 0.001 & 1.01 $\pm$ 0.11 & 1.94 $\pm$ 0.16 & $> 2.0$ & 1.02 & 0.432 \\
Egress-Split (t) & 4 & 353 $\pm 178 $ & 0.995 $\pm$ 0.002 & 0.86 $\pm$ 0.04 & 1.52 $\pm$ 0.04 & $< 2.7$ & 1.09 & 0.168 \\
Egress-Split (t) & 5 & $> 63$ & $> 0.906$ & 1.24 $\pm 0.08$ & $> -0.4 $ & $< 3.2 $ & 3.52 & 1.38$\times 10^{-14}$ \\
\bottomrule
\end{tabular}
\caption{\label{tb:spec} Parameter values and corresponding $1~\sigma$ errors for the spectral fits using model 1 (top, \texttt{tbabs(diskbb+bbodyrad)}, model 2 (middle, \texttt{tbabs(zxipcf * pexriv)} and model 3 (bottom, \texttt{tbabs(zxipcf * xillverns)} to the X-ray bursts observed to peak during totality. Here, we perform fits to the peak of the in-eclipse X-ray bursts and to the peaks (p) and tails (t) of the egress-split X-ray bursts. The bursts are fit simultaneously across \textit{RXTE} gain epochs.}
\end{table*}

\end{appendices}


\bsp 
\label{lastpage}
\end{document}